\documentclass[%
superscriptaddress,
preprint,
 amsmath,amssymb,
 aps,
pra,
]{revtex4-2}

\usepackage{graphicx}
\usepackage{dcolumn}
\usepackage{bm}
\usepackage{xcolor}
\usepackage{multibib}
\newcites{main}{}
\newcites{supp}{Supplementary References}
\begin{document}


\title{Fragmentation and correlations in a rotating Bose-Einstein condensate undergoing breakup}

\author{Sunayana Dutta}
\email{sdutta@campus.haifa.ac.il}
\affiliation{Department of Physics, University of Haifa, Haifa 3498838, Israel}
\affiliation{Haifa Research Center for Theoretical Physics and Astrophysics, University of Haifa, Haifa 3498838, Israel}
\author{Axel U. J. Lode}
\affiliation{Institute of Physics, Albert-Ludwig University of Freiburg, Hermann-Herder-Strasse 3, 79104 Freiburg, Germany}
\author{Ofir E. Alon}
\affiliation{Department of Physics, University of Haifa, Haifa 3498838, Israel}
\affiliation{Haifa Research Center for Theoretical Physics and Astrophysics, University of Haifa, Haifa 3498838, Israel}

\date{\today}

\begin{abstract}
The theoretical investigation of rotating Bose-Einstein condensates has mainly focused on the emergence of quantum vortex states and the condensed properties of such systems.
In the present work, we concentrate on other facets by examining the impact of rotation on the ground state of weakly interacting bosons confined in anharmonic potentials computed both at the mean-field level and particularly at the many-body level of theory.
For the many-body computations, we employ the well-established many-body method known as the multiconfigurational time-dependent Hartree method for bosons (MCTDHB).
We present how various degrees of fragmentation can be generated following the breakup of the ground state densities in anharmonic traps without ramping up a potential barrier for strong rotations.
The breakup of the densities is found to be associated with the acquisition of angular momentum in the condensate due to the rotation.
In addition to fragmentation, the presence of many-body correlations is examined by computing the variances of the many-particle position and momentum operators.
For strong rotations, the many-body variances become smaller than their mean-field counterparts, and one even finds a scenario with opposite anisotropies of the mean-field and many-body variances.
Further, it is observed that for higher discrete symmetric systems of order $k$, namely three-fold and four-fold symmetry, breakup to $k$ sub-clouds and emergence of $k$-fold fragmentation take place.
All in all, we provide a thorough many-body investigation of how and which correlations build up when a trapped Bose-Einstein condensate breaks up under rotation.
\end{abstract}

\maketitle                             


\section{Introduction}
The successful experimental realization of a rotating Bose-Einstein condensate (BEC) has paved the way to explore 
various rich physics of correlated quantum systems \cite{madison2000vortex,raman2001vortex,matthews1999vortices,haljan2001use,abo2002formation}.
Butts and Rokhsar \cite{butts1999predicted} 
first evaluated the wave function of a rotating
BEC using the lowest Landau
level approximation with the help of a Gross-Pitaevskii functional. 
The rotating ultracold  bosonic gases have led to the investigation of the occurrence of quantized vortices \cite{butts1999predicted,matthews1999vortices,madison2000vortex}, vortex nucleation \cite{dagnino2009vortex}, emergence of quantum fluctuations \cite{coddington2003observation,schweikhard2004rapidly},
and presence of the fractional quantum Hall effect in weakly interacting quantum systems \cite{regnault2003quantum}.

There have been studies where breaking up of fast rotating objects is widely detected in systems extending from astronomical objects, e.g., galaxies and supermassive rotating stars \cite{chiappini2011imprints} to the quantum systems in nuclear physics \cite{bohr1979physics}. In atomic physics, the emergence of superfluid flow of a rotating quantum gas has been explored experimentally in Ref. \cite{guo2020supersonic}, in an anharmonic potential. Theoretically, the emergence of breakup in a rotating dipolar condensate was investigated in three-dimension in Ref. \cite{kumar2016three} and finally, in a rotating pancake-like asymmetric quartic-quadratic potential in Ref. \cite{brito2020breakup}. 
 A substantial volume of literature exists corresponding to the rotating BEC, investigating various rich quantum features by employing Gross-Pitaevskii mean-field approximation \cite{aftalion2007vortices,fetter2009rotating,tsatsos2016quantum}. 
However, there exists much less studies of rotating BEC
in the many-body domain, see \cite{viefers2000bose,reimann2006rotating,cremon2013vortices,cremon2015rotating,beinke2018enhanced,khanore2022quantum}, that explore interesting many-body quantum features like the fragmentation of the condensate and correlations.

Condensation and fragmentation are widely explored many-body features of BEC derived from the properties of the one-body reduced density matrix \cite{lowdin1955quantum,yang1962concept,davidson2012reduced,coleman2000reduced,fischer2015condensate,lode2016multiconfigurational}. According to Penrose and Onsager, interacting bosons are said to be condensed if they have a single macroscopic eigenvalue
of the one-body reduced density matrix \cite{penrose1956bose}, and fragmented if there exist two or more
macroscopic eigenvalues \cite{mueller2006fragmentation}.
The fragmentation of condensates has been thoroughly studied for non-rotating systems  \cite{girardeau1962simple,spekkens1999spatial,streltsov2004ground,streltsov2005properties,alon2005pathway,bader2009fragmented,fischer2010interacting,zhou2013fate,kawaguchi2014goldstone,song2014fragmentation,kang2014revealing,jen2015fragmented,fischer2015photonic,fischer2015condensate,lode2016multiconfigurational,kolovsky2017bogoliubov,tomchenko2020fragmented}. However, these many-body features have relatively been less extensively investigated in the rotating frame \cite{sakmann2016single,dagnino2009vortex}. 

In the regime of ``ultrafast rotation'', when the rotation frequency comes closer to the trapping frequency in case of a harmonic trapping potential, the system would tend to escape as the centrifugal force would cancel the trapping force. This problem can be resolved either by introducing an anharmonic term into the confining potential or by adding anisotropy in the harmonic potential. Anisotropy makes the confining potential elongated. Ref. \cite{PhysRevA.72.043613} investigated the appearance of vortex rows of BEC in an anisotropic potential under rotation using Gross-Pitaevskii formalism.
Motivated also by these findings, in this work, we study a system of bosons subject to rotation confined in various 
anharmonic trap geometries with discrete rotational symmetry in two-dimension 2D, namely an elongated trap, a three-fold symmetric trap and finally four-fold symmetric trap. 

There exist several ideas to induce mechanical rotation and coupling of the internal states in the condensate, such as, phase imprinting (using an electromagnetic field) as proposed by Williams and
Holland in Ref. \cite{williams1999preparing}
and optical spoon stirring method where quantized vortices are observed in a stirred gaseous condensate of atomic rubidium \cite{madison2000vortex}. 
Here, we study the physics of a 2D weakly interacting ultracold bosons confined in different anharmonic potentials both at the mean-field level and particularly at the many-body level of theory. Specifically, we investigate how the system transformed from a fully-condensed state into a fragmented state followed by breaking up of the condensate density induced by the rotational motion. As variance is a sensitive probe that characterizes many-body correlations even for fully condensed systems \cite{beinke:15,alon2019analysis}, hence we also analyze the variances of many-particle operators of the fragmented state in the rotating frame. The qualitative difference between the mean-field and many-body variances is a useful tool to explain the nature of many-body correlations.
The many-particle position variance characterizes to what extent a wave-packet spreads or narrows down, similarly the momentum variance is associated with the size of the wave-packet in momentum space. 
Hence, we emphasize on the emergence of correlations by investigating variances of many-particle operators like the position, momentum and angular momentum. Interestingly, the fluctuations present in the system are hardly observed in the angular momentum variance. Thus, we only present the position and momentum variances in the main text and give the detail discussion of the angular momentum variance in the supplemental material.

The time-dependent Gross-Pitaevskii mean-field theory \cite{gross1961structure,pitaevskii1961vortex}
is the most celebrated theoretical model to investigate the many-particle systems of ultracold bosonic atoms. However, this method is unable to study fragmentation and correlations owing to its building via the mean-field ansatz. 
In this paper, we employ a well established many-body numerical method named the multiconfigurational time-dependent Hartree method
for bosons (MCTDHB) \cite{streltsov:07,alon:08} to accurately solve the Schr\"{o}dinger equation at the many-body level  
for ultracold atoms subject to a rotation. The MCTDHB method is a bosonic version of the MCTDH family
of methods \cite{beck:00,wang:03,manthe:08,wang:15,manthe:17,manthe:17_review,bhowmik2020impact,bhowmik2022longitudinal,zanghellini:03,schmelcher:13,schmelcher:17,haxton.pra2:15,miyagi:17,leveque:17,leveque:18,lode:20} which is able to
self-consistently describe the physics involving the presence of many-body correlations. The main focus of the applications of MCTDHB has been the emergence of fragmentation of the condensate,
where the one-body reduced density matrix has multiple significant eigenvalues. For numerical simulation of the results presented in this work, we use the MCTDH-X software \cite{lode:20,lode2019mctdh,lin2020mctdh,lode2021mctdh}.  Finally, the supplemental material reports the benchmarking of MCTDHB for an exactly solvable many-body model under rotation and also presents the convergence of the many-body results of our present work.

\section{Setup and theoretical tools }\label{sec:HAM}
 We consider a system of weakly interacting bosonic atoms in two spatial dimensions 2D confined in non-spherically symmetric trapping potentials in the rotating frame. The properties of these trapped bosons can be described by 
 the (time-dependent) many-body
Schr\"{o}dinger equation. The Schr\"{o}dinger equation dealing with a many-boson system is usually solved by employing the mean-field Gross-Pitaevskii approximation. However, the reduced density matrix involved in the Gross-Pitaevskii approximation has only a single eigenvalue and it involves a single basis state and thereby is unable to capture the many-body features such as fragmentation and correlations. 

In MCTDHB, the (time-dependent) optimized one-body basis is used. Here the basis set and the expansion coefficients in the basis are optimized variationally \cite{streltsov:07,alon:08}. The MCTDHB is a numerically exact method \cite{lode2012numerically} and can describe both coherent and fragmented condensates. MCTDHB includes the theory of Gross-Pitaevskii approximation as a special case
when only a single one-body state is considered. 

\subsection{Hamiltonian}\label{hamil}
The general Hamiltonian of $N$ interacting bosons is given as
\begin{eqnarray}
\hat{H} = \sum_{j=1}^{N} \hat{h}(\mathbf{r}_j) + \sum_{j<k} \hat{W}(\mathbf{r}_j-\mathbf{r}_k), 
\label{eq:hamiltonian}
\end{eqnarray}
where the single-particle Hamiltonian
\begin{eqnarray}
\hat{h}({\bf r }) =  \hat{T}(\mathbf{r}) + \hat{V}(\mathbf{r})\label{single}
\end{eqnarray}
is composed of the
kinetic energy and the external potential energy, respectively. Here, the 
interaction of ultracold dilute bosonic gases is considered to be a finite range interaction and modelled by a Gaussian function \cite{doganov2013two,bhowmik2020impact,bhowmik2022longitudinal}, $\hat{W}(\mathbf{r}-\mathbf{r}')=\frac{\lambda_0}{2 \pi \sigma^2} e ^{-\frac{({\bf r}-{\bf r'})^2}{2\sigma^2}}$ with $\sigma=0.25$. This avoids the regularization of the delta contact potential in 2D. The interaction strength $\lambda_0$ is scaled with the number of bosons $N$ as $\Lambda=\lambda_0(N-1)$, where $\Lambda$ is the interaction parameter. One uses the interaction parameter to define the mean-field regime.
In our study, we work in the units $\hbar=m=1$ and all the quantities are dimensionless. We also consider three different trapping potentials $\hat{V}({\bf r})$ that we shall discuss in the next section. The first setup is the elongated trap $-$ that leads to breaking up of the ground state density into two clouds. After that we move to more complex traps, namely three-fold symmetric and four-fold symmetric traps to investigate and establish the generality of the results. Hence, our strategy is to first study an elongated trap, and then a three-fold symmetric trap, and finally, a four-fold symmetric trap to see what stays between the two to three and the three to four-fold symmetric traps.

In the rotating frame, the kinetic energy operator is modified and can be written as,
\begin{eqnarray}
\hat{T}({\bf r })  = \frac{1}{2} (\hat{p}_{x}^2+\hat{p}_{y}^2) - \omega_r\hat{l}_z,\label{rotation}
\end{eqnarray}
here $\omega_r$ is the rotation frequency and $\hat{l}_z=\hat{x}\hat{p}_y-\hat{y}\hat{p}_x$ is the angular-momentum operator.

An alternative way to mimic the rotational effect in the condensate is by introducing a synthetic gauge field $\mathbf{A}(\mathbf{r})$ as
\begin{equation}
\hat{T}(\mathbf{r})= \frac{1}{2} \big[-i\nabla_{\mathbf{r}} - q \mathbf{A}(\mathbf{r}) \big]^2.\label{arti_kin}
\end{equation}
Consider the following general form of the gauge field: 
\begin{equation}
\mathbf{A}(\mathbf{r})=(ay,bx,0).
\end{equation}
Then, expansion of Equation \eqref{arti_kin} leads to,
\begin{eqnarray}
 \hat{T}(\mathbf{r})= \frac{1}{2} (\hat{p}_{x}-ay)^2+\frac{1}{2}(\hat{p}_{y}-bx)^2 
                 = \frac{1}{2} (\hat{p}_{x}^2+\hat{p}_{y}^2) -  (\hat{p}_x ay+\hat{p}_y bx)+\frac{1}{2}(a^2y^2+b^2x^2).
 \label{eq:kinetic_gauge}
\end{eqnarray}
For the specific case $b=-a$, Equation \eqref{eq:kinetic_gauge} becomes
\begin{equation}
\hat{T}(\mathbf{r})= \frac{1}{2} (\hat{p}_{x}^2+\hat{p}_{y}^2) - a\hat{l}_z + \frac{1}{2}a^2(y^2+x^2).
\label{eq:kinetic_gauge_rotation}
\end{equation}
Combining Equations \eqref{single}, \eqref{rotation} and \eqref{eq:kinetic_gauge_rotation} we have
\begin{equation}
 \hat{h}({\bf r })= \hat{T}(\mathbf{r}) + \hat{V}'(\mathbf{r}),
\end{equation}
where the modified confining potential is $\hat{V}'(\mathbf{r})=\hat{V}(\mathbf{r})+\frac{1}{2}a^2\mathbf{r}^2$ and $a=\omega_r$ corresponds to the rotation frequency of the condensate.
\subsection{Many-body method}
The MCTDHB method employs time-adaptive orbitals to represent the field operator as a sum of the $M$ time-dependent single-particle states 
\begin{equation}
\hat{\Psi}(\mathbf{r},t)= \sum_{j=1}^{M} \hat{b}_j \phi_j(\mathbf{r},t).
\end{equation}
The ansatz of the MCTDHB wavefunction is 
\begin{equation}
\vert \Psi(t) \rangle = \sum_{\vec{n}} C_{\vec{n}} \vert \vec{n},t \rangle. \label{eq:ansatz}
\end{equation}
The summation in Equation \eqref{eq:ansatz} runs over all  
$\binom{N+M-1}{N}$
possible time-dependent configurations $\vec{n}=(n_1,...,n_M)$ with fixed particle number $N=\sum_{i=1}^{M} n_i$. 
To derive the MCTDHB equations, the time-dependent variational principle~\cite{dirac:30,frenkel:34,kramer:81} is employed for the ansatz in Equation~\eqref{eq:ansatz}. Thus, resulting in two-coupled equations of motion -- a set of linear equations for the coefficients $\lbrace C_{\vec{n}} \rbrace$ and a set of non-linear equations for the orbitals $\lbrace \phi_j(\mathbf{r}); j=1,...,M\rbrace$, see Refs.~\cite{alon:08,streltsov:07} for details and derivation of the equations of motion. 
In the following work, the self-consistent ground state is achieved by relaxing the system via imaginary-time propagation and is hence determined by the variational principle. Thus, in the following sections, we omit the time-dependency from the various quantities and observables that are involved in the many-body simulations.
\subsection{Quantities of interest}\label{sec:QOI}
In this section, we define the quantities of interest, namely the one-body density, the eigenvalues of the one-body reduced density matrix (RDM), the expectation value of the angular momentum operator, and finally, the many-particle variances of the position, momentum, and angular momentum operators.

\subsubsection{One-body reduced density matrix (RDM), one-body density, and natural occupations}
The one-body RDM of the N-boson state $|\Psi\rangle$ is a hermitian matrix and is defined as
\begin{equation}
\rho^{(1)}(\mathbf{r},\mathbf{r}') = \langle \Psi \vert\hat{\Psi}^{\dagger}(\mathbf{r}') \hat{\Psi}(\mathbf{r})\vert  \Psi \rangle = \sum_{k,q} \rho_{kq} \phi^*_k(\mathbf{r}') \phi_q(\mathbf{r})\label{eq:1BDM}
\end{equation}
in its eigenbasis $\{\phi_q(\mathbf{r})\}$.
The matrix elements $\rho_{kq}=\langle \Psi \vert \hat{b}^\dagger_k \hat{b}_q \vert \Psi \rangle$ represent the one-body RDM using $M$ orbitals corresponding to the creation (annihilation) operators $\hat{b}^\dagger_k$ ($\hat{b}_q$).
The diagonal of $\rho^{(1)}(\bf{r},\bf{r}')$ is referred to as the one-body density $\rho(\mathbf{r})$ which is
$\rho(\mathbf{r}) = \rho^{(1)}(\mathbf{r},\mathbf{r}'=\mathbf{r})$.

The eigenvalues of the one-body RDM are obtained by the diagonalization of Equation~\eqref{eq:1BDM} which corresponds to a unitary transformation of the orbitals $\phi_q(\mathbf{r}) $ to the natural orbitals $\phi_j^{(NO)}(\mathbf{r})$ as 
\begin{equation}
\frac{\rho^{(1)}(\mathbf{r},\mathbf{r}')}{N} = \sum_j n_j \phi_j^{(NO),*}(\mathbf{r}') \phi_j^{(NO)}(\mathbf{r}).
\end{equation}
Here, the eigenvalues $n_j$ are normalized as $\sum_{j=1}^M n_j=1$ and, without loss of generality, they are sorted in magnitude such that $n_1\geq n_2\geq...$ throughout this work.  
The eigenvalues $n_j$ are termed natural occupations and characterize the degree of condensation and fragmentation of the bosons. Thus, the system with one-body RDM consisting only a single macroscopically-contributing eigenvalue $n_1$ is said to be condensed~\cite{penrose1956bose}. When the one-body RDM has $k$ macroscopically-occupied eigenvalues, the system is referred to as $k$-fold fragmented~\cite{spekkens1999spatial}.

\subsubsection{Angular momentum}
For a 2D many-particle systems, there is only a single component of the angular momentum operator, i.e.,
\begin{eqnarray}
\hat{L}_Z = \sum_{j=1}^N\hat{l}_{z_j}=\sum^N_{j=1}\frac{1}{i} \left( \hat{x}_j \frac{\partial}{\partial y_j} - \hat{y}_j\frac{\partial}{\partial x_j} \right).\label{eq:AM}
\end{eqnarray}
Bosonic systems with angular momentum provide rich quantum features beyond mean-field theory, such as, phantom vortices~\cite{weiner:17}, spatially partitioned  vortices~\cite{beinke:15}, and fragmentation~\cite{tsatsos:10,beinke:15,weiner:17}. 
In the following studies, we investigate the expectation value of the angular momentum operator per particle $\frac{1}{N}\langle \Psi|\hat{L}_Z|\Psi\rangle$, for three different confining traps to see the effect of the rotation. In our system, we expect an intricate dynamics of the angular momentum acquisition under rotation and build up of correlations.  
\subsubsection{Many-particle variances}
The variance of a many-particle observable $\hat{O}=\sum_{j=1}^N\hat{o}(\mathbf{r}_j)$ can be written as \cite{klaiman2015variance}
\begin{eqnarray}
\frac{1}{N}\Delta^2_{\hat{O}}&=&\frac{1}{N}\bigg[\langle\Psi|\hat{O}^2|\Psi\rangle-\langle\Psi|\hat{O}|\Psi\rangle^2\bigg] =\nonumber\\
                             &=&\frac{1}{N} \bigg\{ \sum_j n_j \int \mathrm{d} \mathbf{r} \: \phi^{*(NO)}_j(\mathbf{r}) \hat{o}^2\phi^{(NO)}_j(\mathbf{r})-\Bigg[\sum_j n_j \int \mathrm{d} \mathbf{r} \: \phi^{*(NO)}_j(\mathbf{r}) \hat{o}\phi^{(NO)}_j(\mathbf{r}) \Bigg]^2+\nonumber\\
                              &+& \sum_{jklm}\rho_{jklm} \bigg[\int \mathrm{d} \mathbf{r} \phi^{*(NO)}_j(\mathbf{r}) \hat{o}\phi^{(NO)}_l(\mathbf{r})\bigg]\bigg[\int \mathrm{d} \mathbf{r} \phi^{*(NO)}_k(\mathbf{r}) \hat{o}\phi^{(NO)}_m(\mathbf{r})\bigg] \bigg\}.\label{variance_formula}
\end{eqnarray}
Here, the expectation value of $\hat{O}=\sum_{j=1}^N\hat{o}(\mathbf{r}_j)$ is dependent only on the one-body operator, whereas the expectation of $\hat{O}^2$ is a combination of one- and two-body operators $\hat{O}^2=\sum_{i=1}^N\hat{o}^2(\mathbf{r}_j)+\sum_{j<k}2\hat{o}(\mathbf{r}_j)\hat{o}(\mathbf{r}_k)$. $\rho_{jklm}$ are the two-particle reduced density matrix elements, $\rho(\mathbf{r}_1,\mathbf{r}_2,\mathbf{r}'_1,\mathbf{r}'_2)=\sum_{jklm}\rho_{jklm} \phi^{*}_j(\mathbf{r})\phi^{*}_k(\mathbf{r})\phi_l(\mathbf{r})\phi_m(\mathbf{r})$. 
In the following work, to analyze the emergence of many-body correlations in the rotating condensate, we study the many-particle variances per particle of the position, momentum, and angular momentum operators \cite{alon2019analysis}.
 
\section{Many-body physics of Bose-Einstein condensates breaking up under rotation: Results and Discussion}
We investigate the impact of rotation on the ground state of weakly interacting bosonic atoms by dividing the analysis into two main parts depending on the confining anharmonic potentials. 
First, the breakup and fragmentation processes in an elongated trap are investigated and then the breakup and fragmentation in more complicated traps of discrete spatial symmetry are explored.
In the following sections, we  analyze the effect of rotation on various static properties, namely the ground state energy, one-body density, natural occupations, expectation value of the angular momentum operator, and finally, the variances of the many-particle position, momentum, and angular momentum operators as a function of the rotation frequency $\omega_r$.
For the numerical computations, the MCTDH-X implementation of the MCTDHB theory is employed \cite{lode:20,lode2019mctdh,lin2020mctdh,lode2021mctdh}. 
In our work, we consider $N=10$ weakly interacting ultracold bosonic atoms interacting via a Gaussian interaction \cite{doganov2013two,bhowmik2020impact,bhowmik2022longitudinal} with the interaction parameter chosen as $\Lambda=0.1$ throughout the computations. 
The grid used to represent the Hamiltonian Equation~\eqref{eq:hamiltonian}, see section Sec.\ref{hamil}, extends from 
$[-8,8)$ to $[-8,8)$ and comprises of $128\times 128$ discrete variable representation (exponential) functions to represent each of the orbitals
${\phi_j(\bf{r})}$. 

 \begin{figure}[htbp]
 \centering
  \includegraphics[angle=270,width=0.45\textwidth]{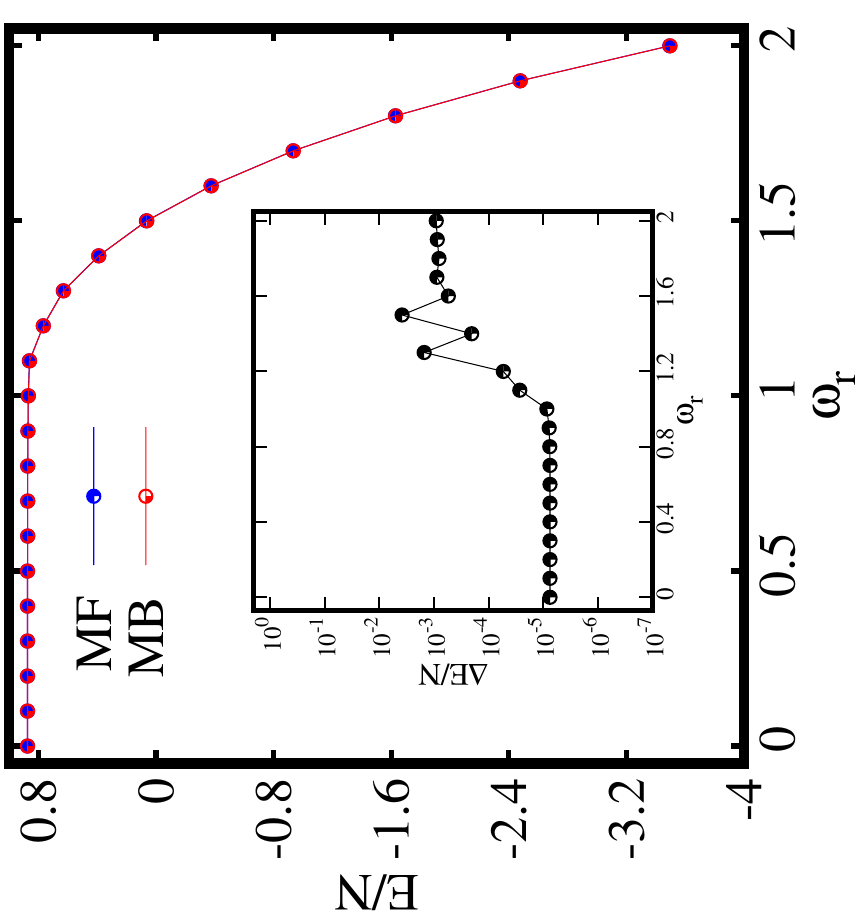}
 \caption{Behaviour of the ground state energy per particle E/N as a function of the rotation frequency $\omega_r$ in an elongated trap computed at the mean-field (MF) and many-body (MB) levels (with M=1 and M=4 self-consistent orbitals, respectively). The inset shows the energy difference between the mean-field and many-body energies. All quantities shown are dimensionless.}\label{energy_elongated}
 \end{figure}
\subsection{Breaking up to two clouds and pathway to two-fold fragmentation}
Let us discuss the impact of rotation on the
weakly interacting bosons trapped in an elongated confinement. The elongated trap is an anharmonic potential that elongates the condensate in the $x$-direction and can be written as
\begin{eqnarray}
 V(\mathbf{r})=\frac{1}{4} (0.8x^2+y^2)^2.
\label{eq:elongated}
\end{eqnarray}
For this trap, we consider $M=4$ self-consistent orbitals to investigate the ground state properties of the condensate. We recomputed these results with $M=8$ self-consistent orbitals to check the convergence of the system, see the elaborate discussion in the supplemental material. The rotation frequency range is $\omega_r=[0,2.0]$ in this trap.

 Figure \ref{energy_elongated} shows the behaviour of the ground-state energy per particle E/N, in the rotating frame computed both at the mean-field and many-body levels. Initially, E/N remains almost constant for slow rotation. Then, E/N drops gradually with further increase in $\omega_r$ as evident from Figure \ref{energy_elongated}. It is also observed that the energies computed both at the mean-field and many-body 
\begin{figure}[htbp]
  \centering
  \includegraphics[angle=270,width=1.0\textwidth]{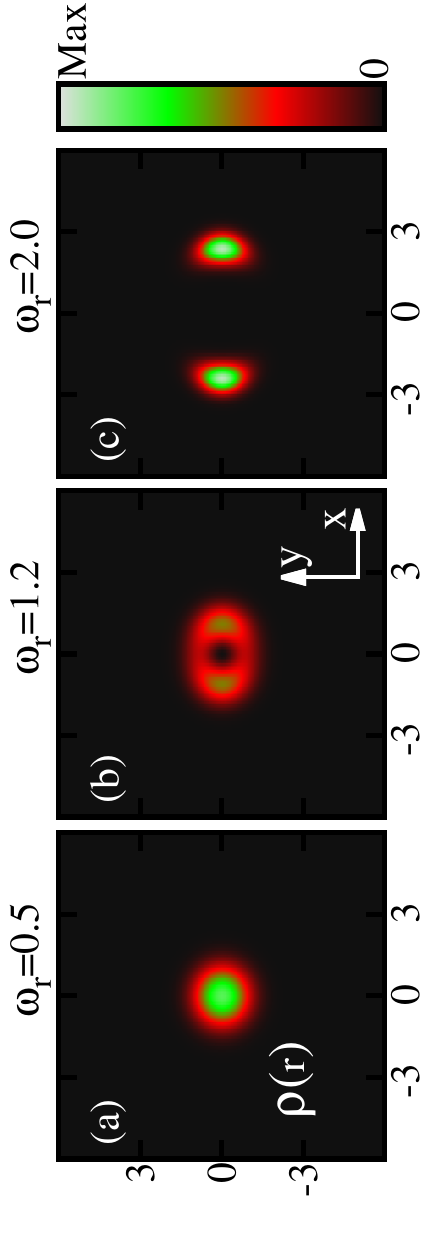}
 \caption{The one-body densities per particle are shown for three different rotation frequencies $\omega_r$ in the  elongated trap at the many-body level. M=4 self-consistent orbitals are used. The density computed at the mean-field (not shown) and many-body levels depict identical features for all $\omega_r$. All quantities shown are dimensionless.}\label{fig:elong}
 \end{figure}
 levels practically coincide with each other for all $\omega_r$. The inset of Figure \ref{energy_elongated} corresponds to the energy difference per particle $\Delta E/N$, between the mean-field and many-body energies defined as 
 $\Delta E =E_{MF}-E_{MB}$. The energy difference remains minimum till about $\omega_r=1.2$. Subsequently, $\Delta E/N$ exhibits some structures for an intermediate range of $\omega_r$ around $\omega_r\sim 1.3$. Finally, the energy difference slightly rises from of the order of $10^{-5}$ to $10^{-3}$ for larger rotation frequencies $\omega_r \ge 1.6$. The presence of these structures at the intermediate rotation might suggest that something interesting is happening at the many-body level. Therefore, we dig deeper than the energy of the system to see the many-body features.

Now let us discuss the behaviour of the ground-state densities per particle $\frac{\rho(\mathbf{r})}{N}$ of the rotating condensate confined in the elongated trap.  
The densities computed at the many-body level of theory are shown for three different rotation frequencies $\omega_r$ in Figure \ref{fig:elong}. 
In absence of rotation, the density displays a single cloud where all bosons accumulate in the center of the trap. This behaviour persists with the inclusion of rotation for slow rotation frequencies, e.g., at $\omega_r=0.5$. 
Further increase in rotation induces a breaking of the density into two clouds. The breakup of the ground-state density confined in the potential given by Equation \eqref{eq:elongated} corresponds to the scenario where the minimum of the potential is splitted into two parts and shifted with rotation by creating an effective double-well potential. 
It is also observed that the distance between the two densities increases with increase of $\omega_r$ [Figure \ref{fig:elong}(b)-(c)]. 
We have also computed the density per particle at the mean-field level and identical features are observed in the density profile. 
Hence, it is observed that the densities computed at the mean-field and many-body levels in the real space show identical pattern. 
In the density profile, we show three specific frequencies that correspond to slow, fast, and faster rotations.

\begin{figure}[htbp]
  \centering
  \includegraphics[angle=270,width=0.45\textwidth]{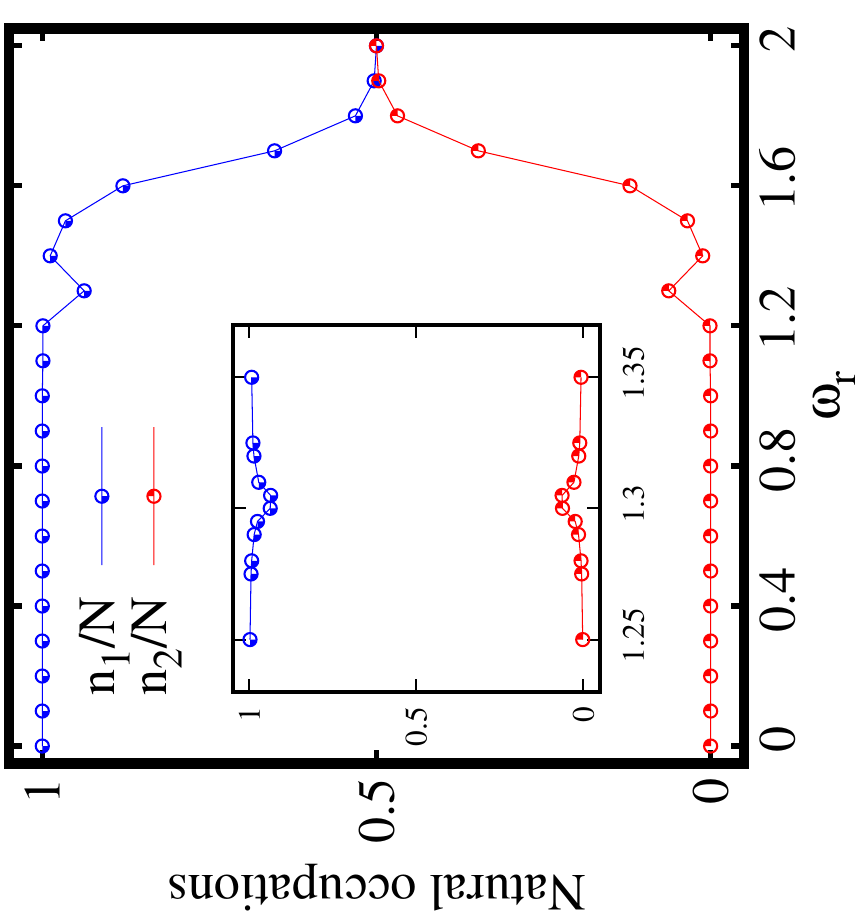}
 \caption{Pathway from condensation to fragmentation in the rotating elongated trap.
 two-fold fragmentation is observed with increase in rotation. 
 The two leading natural occupations $n_1/N$ and $n_2/N$, are shown as a function of the rotation frequency $\omega_r$. $M=4$ self-consistent orbitals are used. 
 The third and fourth natural occupations satisfy $n_3/N$, $n_4/N\le 10^{-5}$, see the supplemental material. All quantities shown are dimensionless.}\label{occupation_elong}
\end{figure}

 To understand whether the splitting of the density into two clouds [as shown in Figure \ref{fig:elong}(c)] for faster rotations leads to fragmentation of the condensate, we further discuss the dependence of the natural occupations $\frac{n_j}{N}$. In the elongated trap, four natural occupations corresponding to the four natural orbitals are employed and found to vary with the rotation frequency $\omega_r$ (see Figure \ref{occupation_elong}).
 The system remains fully condensed, i.e., $\frac{n_1}{N}\sim1$, $\frac{n_2}{N}\sim \frac{n_3}{N}\sim \frac{n_4}{N}\le 10^{-5}$ with the inclusion of rotation till about $\omega_r=1.1$ [see Figure S2(a) of the supplemental material, which displays the depletion as a function $\omega_r$ on a log scale]. As $\omega_r$ increases further, the first natural occupation number $n_1/N$, falls of gradually following an increase in population of the second natural occupation number $n_2/N$. 
 The other two natural occupation numbers remain almost the same as $\frac{n_3}{N}\sim \frac{n_4}{N}\le 10^{-5}$.
For faster rotation at $\omega_r=2$, the state becomes essentially fully two-fold fragmented with natural occupations of $\frac{n_1}{N}\approx \frac{n_2}{N} \approx 50\%$. 
This signifies equally populated two leading natural orbitals, whereas $\frac{n_3}{N}$ and $\frac{n_4}{N}$ remain essentially unpopulated. 
For an intermediate frequency, $\omega_r=1.3$, we observe a scenario where $n_1$ shows a deep followed by a peak in $n_2$. 
To understand this feature, we zoom in at the intermediate points between the rotation frequencies $\omega_r=1.2$ and $\omega_r=1.4$ as shown in the inset of Figure \ref{occupation_elong}. 
A smooth transition is found, from coherence via loss of coherence to build up of coherence. 
This transition might suggest the presence of a resonant-like behaviour of the interacting bosons in the elongated 2D trap at this specific rotation frequency.

To intermediately summarize, for weakly interacting bosons confined in an elongated trap [Equation \eqref{eq:elongated}], inclusion of rotation triggers a transition from a fully condensed state to a fully two-fold fragmented state with equally populated two leading natural orbitals. 
Thus, it can be concluded that the rotation can be used as a tool to manipulate fragmentation.

 \begin{figure}[htbp]
 \centering
  \includegraphics[angle=270,width=0.45\textwidth]{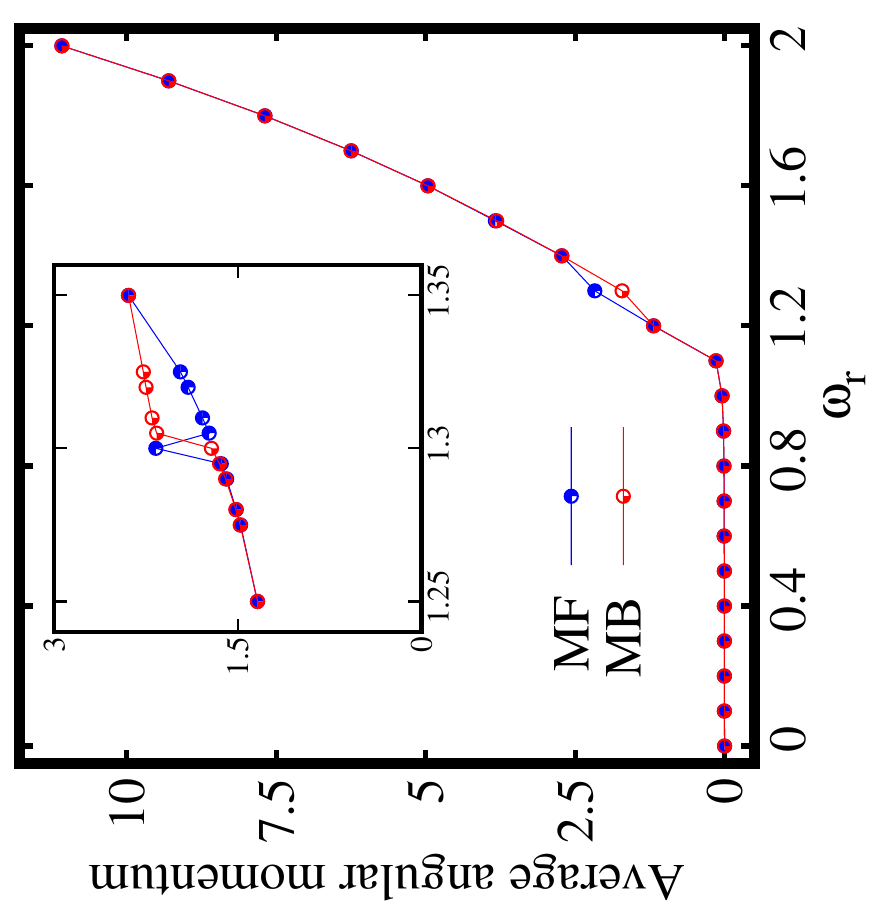}
 \caption{Expectation value of angular momentum operator $\langle \hat{L}_Z \rangle/N$, computed at the mean-field (MF) and many-body (MB) levels [with $M=1$ and $M=4$ self-consistent orbitals, respectively] as a function of the rotation frequency $\omega_r$ for the elongated trap. Actual data are points. The continuous curves are only to guide the eye. All quantities shown are dimensionless.}\label{angular_momentum_1}
 \end{figure}
Next, let us discuss the behaviour of the expectation value of the angular momentum per particle $\frac{\langle\psi|\hat{L}_Z|\psi\rangle}{N}$, computed at the mean-field and many-body levels as a function of the rotation frequency $\omega_r$. 
It is observed from Figure \ref{angular_momentum_1} that the mean-field and many-body angular momenta exactly coincide each other at all $\omega_r$ except for the intermediate rotation frequency around $\omega_r=1.3$ [see the inset of Figure \ref{angular_momentum_1}].
 The angular momentum remains minimum till about $\omega_r=1.1$. 
 For $\omega_r=1.2$ (the rotation frequency that corresponds to the breakup of the density), the rotation generates a state where significant angular momentum enters the system with $\langle \hat{L}_Z\rangle/N > 1$. 
 The angular momentum of the condensate gradually increases with further increase of the rotation. 
 
 As we know, the variance is a sensitive probe of correlations that allows one to study the quantum fluctuations present in a system \cite{beinke:15,alon2019analysis}. 
 Thus, it would be interesting to investigate the variance of many-particle operators which signifies many-body correlations for the fragmented condensate in the rotating frame. 
 Therefore, we further analyze the impact of rotation on the behaviour of the many-particle variances of position and momentum operators which are sensitive to rotation. 
 \begin{figure}[!t]
  \centering
  \includegraphics[angle=270,width=0.45\textwidth]{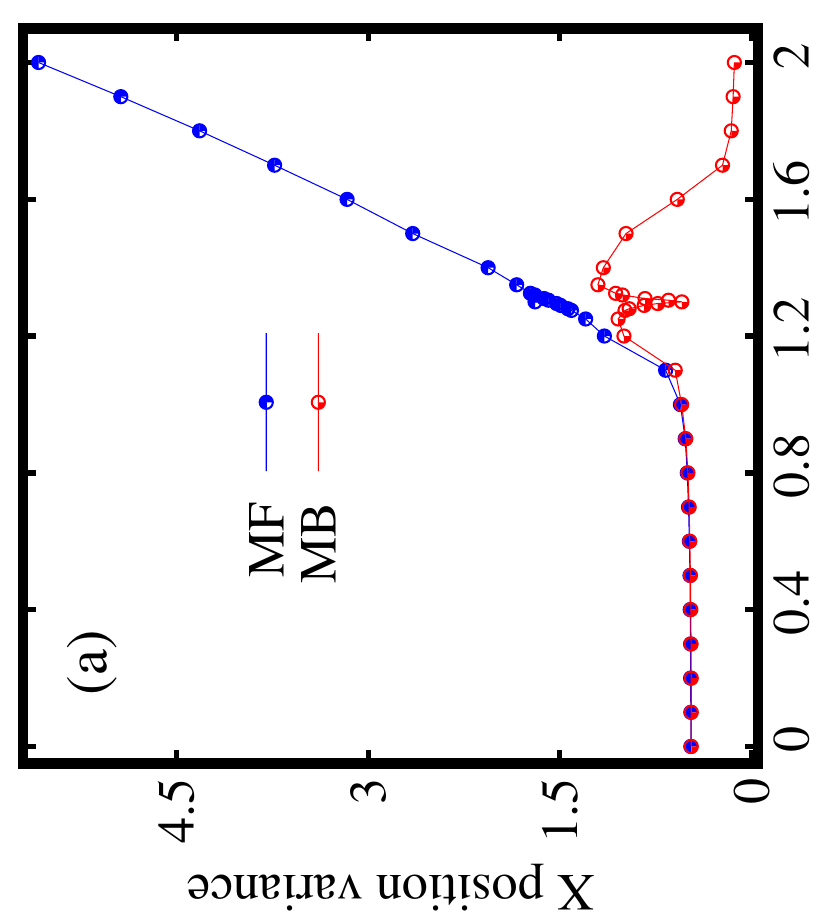}
  \includegraphics[angle=270,width=0.45\textwidth]{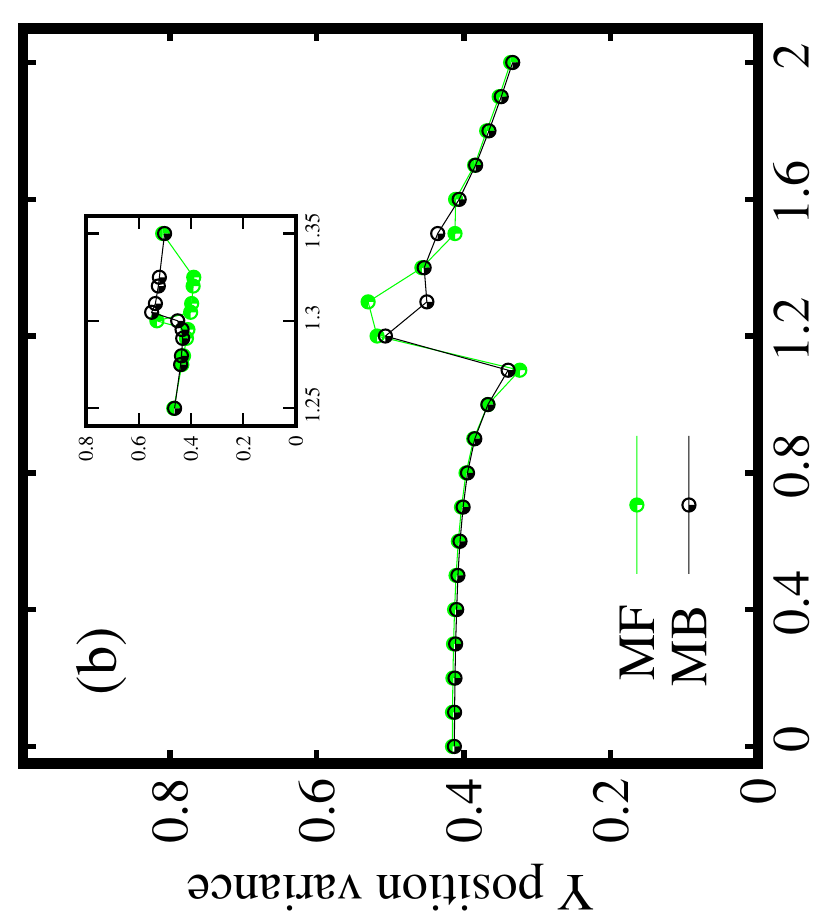}\\
  \includegraphics[angle=270,width=0.45\textwidth]{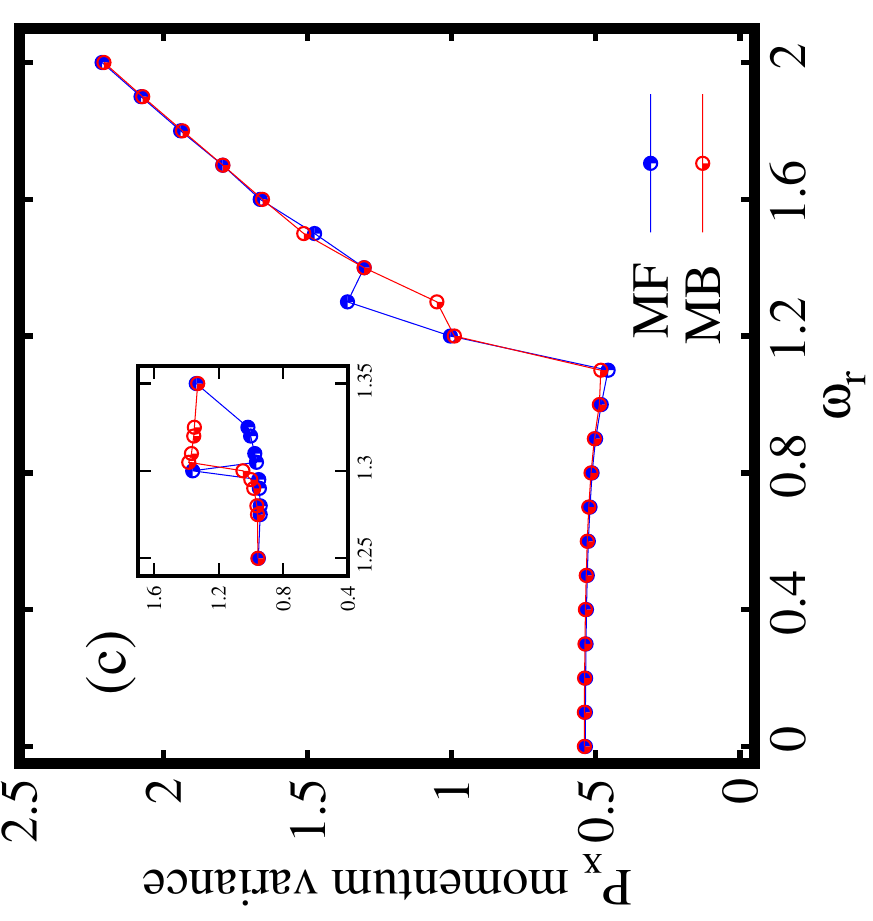}
  \includegraphics[angle=270,width=0.45\textwidth]{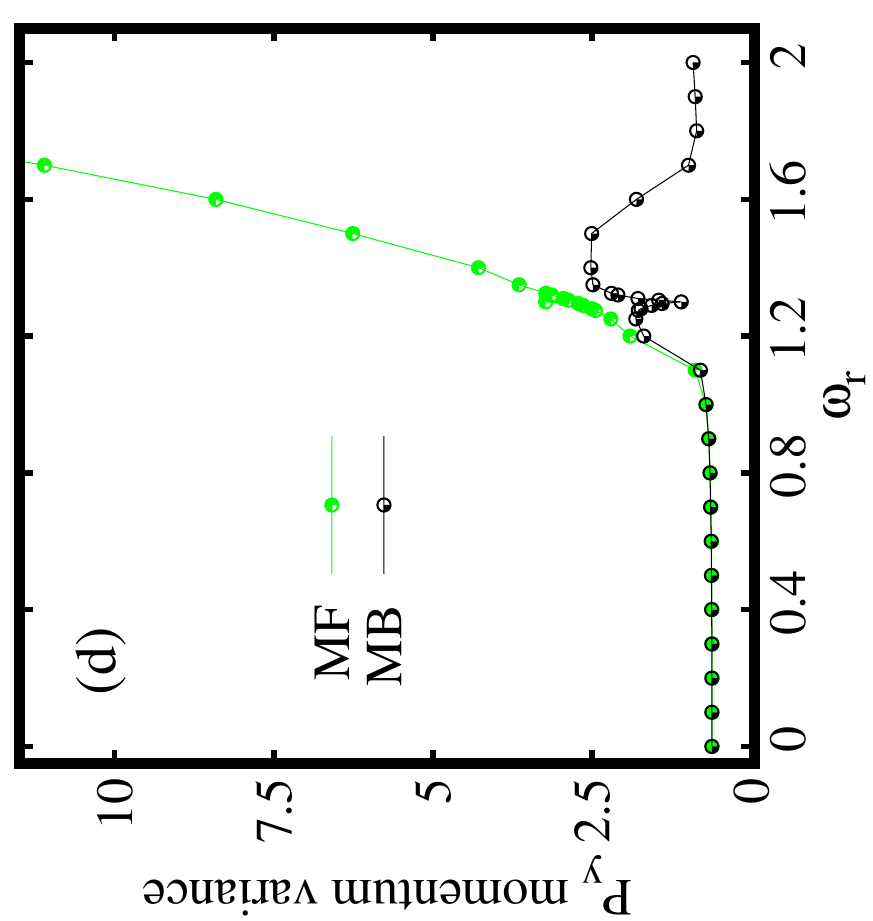}
 \caption{Dependence of the the many-particle position and momentum variances on the rotation in the elongated trap. Shown are (a) $\frac{1}{N}\Delta^2_{\hat{X}}$, (b) $\frac{1}{N}\Delta^2_{\hat{Y}}$, (c) $\frac{1}{N}\Delta^2_{\hat{P}_X}$, and (d) $\frac{1}{N}\Delta^2_{\hat{P}_Y}$ as a function of $\omega_r$ at the many-body level (MB) [$M=4$ self-consistent orbitals] and at the mean-field level (MF) [$M=1$ self-consistent orbital]. All quantities shown are dimensionless.}\label{variance_position_elong}
 \end{figure}
 
 Figures \ref{variance_position_elong}(a)-(b) display the behaviour of the many-particle position variance per particle $\frac{1}{N}\Delta^2_{\hat{X},\hat{Y}}$ along the $x$ and $y$-directions respectively, as a function of the rotation frequency $\omega_r$ in the elongated trap. 
 The mean-field and many-body position variances along the $x$- direction $\frac{1}{N}\Delta^2_{\hat{X}}$, coincide till about $\omega_r=0.9$, indicating the essential absence of correlations in the system [Figure \ref{variance_position_elong}(a)]. 
 Further, the mean-field position variance monotonically increases with $\omega_r$ which signifies spreading of the density as observed in Figure \ref{fig:elong}. 
 Also, the many-body position variance slowly increases till $\omega_r=1.2$. 
 But, now, we observe a deep at $\omega_r=1.3$ that corresponds to the loss of coherence in the natural occupation, see Figure \ref{occupation_elong}. 
 After increasing and reaching a maximum at rotation frequency of about $\omega_r\sim 1.4$, $\frac{1}{N}\Delta^2_{\hat{X}}$ starts to decrease which incorporates the emergence of a small amount of depletion of the condensate [see Figure \ref{variance_position_elong}(a)]. 
 For a faster rotation, at $\omega_r=2$, the position variance $\frac{1}{N}\Delta^2_{\hat{X}}$ decreases significantly which goes hand in hand with the emergence of fragmentation of the condensate. 
 Similar behaviour of the position variance along the $x$-direction is observed in a two-dimensional double well in \cite{alon2019variance}, albeit without the resonant-like behaviour described above.
 The position variance per particle along the $y$-direction $\frac{1}{N}\Delta^2_{\hat{Y}}$ is almost frozen and varies slowly as shown in Figure \ref{variance_position_elong}(b). 
 This might suggest that excitations along the tighter $y$-directions are practically not involved, at least as far as the position variance is considered. 
 The $\frac{1}{N}\Delta^2_{\hat{Y}}$ computed at the mean-field level  matches that at the many-body level almost for all the rotation frequencies except of small difference for the intermediate $\omega_r$. 
 As above, convergence of the results is detailed in the supplemental material.
 
 Further, we can describe the anisotropy of the variance by considering two facts, \\
 $(1)$ by comparing two quantities, one along the $x$- direction and the other along the $y$- direction, whether they are similar or different.\\
 $(2)$ by comparing these quantities at the mean-field and many-body levels which demonstrates whether the anisotropy of the many-particle variances are alike or opposite.   
 Hence, in the elongated trap and for small rotations, it is observed that
 \begin{center}
 $\frac{1}{N}\Delta^2_{\hat{X}}|_{MB}<\frac{1}{N}\Delta^2_{\hat{Y}}|_{MB}$, at MB\\
  $\frac{1}{N}\Delta^2_{\hat{X}}|_{MF}<\frac{1}{N}\Delta^2_{\hat{Y}}|_{MF}$, at MF.
 \end{center}
  Thus, the anisotropy of the many-particle position variance computed at the mean-field and many-body levels are alike at slow rotation.   
 However, for fast rotation we find
 \begin{center}
 $\frac{1}{N}\Delta^2_{\hat{X}}|_{MB}<\frac{1}{N}\Delta^2_{\hat{Y}}|_{MB}$, at MB\\
  $\frac{1}{N}\Delta^2_{\hat{X}}|_{MF}>\frac{1}{N}\Delta^2_{\hat{Y}}|_{MF}$, at MF.
 \end{center}
 Hence, it indicates that the many-particle position variances display opposite anisotropy when computed at the mean-field and many-body levels.
 
 Now, we investigate the behaviour of the many-particle momentum variance per particle $\frac{1}{N}\Delta^2_{\hat{P}_X,\hat{P}_Y}$ along the $x$- and $y$-directions as a function of rotation frequency $\omega_r$ in the elongated trap [Figure \ref{variance_position_elong}(c)-(d)]. It is observed that
 unlike the position variance, the mean-field and many-body momentum variances along the $x$-direction are almost similar, see Figure \ref{variance_position_elong}(c). The momentum variances computed at the  mean-field and many-body levels gradually increase from $\omega_r=1.2$ onwards. This corroborates the narrowing of the density lobes along the $x$-direction in real space.
The momentum variance in the $y$-direction displays a completely different picture, see Figure \ref{variance_position_elong}(d). The mean-field and many-body variances match each other till $\omega_r=1.1$. However, for a faster rotation, the momentum variance computed at the mean-field deviates from that computed at the many-body level in the $y$-direction. $\frac{1}{N}\Delta^2_{\hat{P}_Y}$ displays similar behaviour as the $\frac{1}{N}\Delta^2_{\hat{X}}$ as clear from Figure \ref{variance_position_elong}(a). Thus, excitations along the $y$-direction plays a role in the momentum space as a result of rotation.

Inverse to the position variance, in the case of the momentum variance
for slow rotation we find 
\begin{center}
 $\frac{1}{N}\Delta^2_{\hat{P}_X}|_{MB}>\frac{1}{N}\Delta^2_{\hat{P}_Y}|_{MB}$, at MB\\
  $\frac{1}{N}\Delta^2_{\hat{P}_X}|_{MF}>\frac{1}{N}\Delta^2_{\hat{P}_Y}|_{MF}$, at MF.
 \end{center}
Hence, the anistropy of the momentum variances are alike when computed at the many-body and mean-field levels of theory for slow rotation. 
 For a faster rotation however, we find
 \begin{center}
 $\frac{1}{N}\Delta^2_{\hat{P}_X}|_{MB}>\frac{1}{N}\Delta^2_{\hat{P}_Y}|_{MB}$, at MB\\
  $\frac{1}{N}\Delta^2_{\hat{P}_X}|_{MF}<\frac{1}{N}\Delta^2_{\hat{P}_Y}|_{MF}$, at MF.
 \end{center}
 This signifies the presence of opposite anisotropy of the momentum variance. This feature was not found before for static double well \cite{alon2019variance}. Hence, this indicates one of the distinct features of the rotation, which makes both the many-particle position and momentum variances show opposite anisotropies with respect to the mean-field and many-body levels in the elongated trap. The rotation localized the position variance in the long direction and the momentum variance in the narrow direction and this is purely a many-body effect.
 
Let us briefly summarize the results so far. 
The ground state of a Bose-Einstein condensate in a two-dimensional elongated trap is analyzed in presence of rotation.
Here, the ground state density splits into two clouds for fast rotation.
It is fascinating to observe the effect of rotation on the condensate as it leads to a transition from condensed state to a fully two-fold fragmented state in a single well without ramping up a barrier. 
Thus, rotation can be used as a probe to manipulate various degrees of fragmentation which we will further discuss in the next section where higher-order fragmentations are explored. 
Further, we observe the presence of an opposite anisotropy both in the position and momentum variances by comparing the mean-field and many-body results for fast rotations.

\subsection{Breaking up to several clouds and pathway to higher-order fragmentation}
So far in the elongated trap, we observe breaking up of the ground state density into two clouds, followed by the emergence of two-fold fragmentation and presence of opposite anisotropy both in the position and momentum variances for faster rotations.
Now, it would be fascinating to explore some more complex potentials with $k$-fold rotational symmetry, explicitly three-fold and four-fold symmetric potentials, to analyze the impact of rotation on generic properties. Would the density break into more than two clouds? Is higher-order fragmentation possible? How would the angular momentum enter into the condensate? Finally, when correlations set in, who wins, the many-body or the mean-field variance? Both for the position and momentum variances?
\subsubsection{Bosons in a three-fold symmetric trap under rotation}
 In this section, we analyze the impact of rotation on weakly interacting bosons confined in a three-fold symmetric trap of the form
\begin{eqnarray}
 V({\bf{r}}) = \frac{1}{5} (x^2+y^2)^2+\frac{1}{5}(x^2y-\frac{1}{3}y^3).
\label{eq:potentials_three}
\end{eqnarray}
In this trap, we consider $M=3$ self-consistent orbitals to compute the ground state properties of the system. 
We also computed the following results with $M=6$ self-consistent orbitals for convergence and checked the consistency of our results (see the supplemental material). 
The rotation frequency range is  $\omega_r=[0,2.5]$ for the following analysis.
\begin{figure}[htbp]
  \centering
  \includegraphics[angle=270,width=1.0\textwidth]{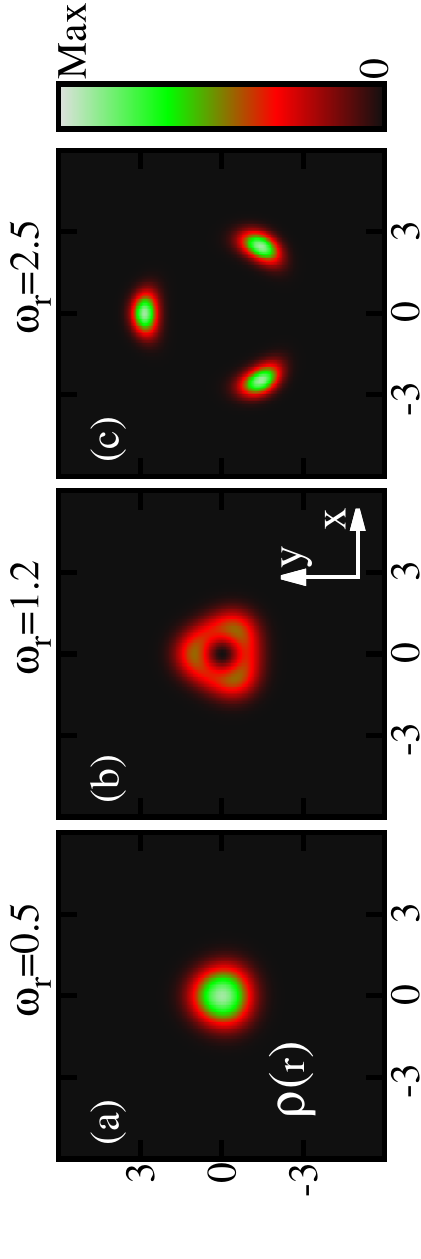}
 \caption{The one-body densities per particle are shown for three different rotation frequencies $\omega_r$ in the three-fold symmetric trap at the many-body level. M=3 self-consistent orbitals are used. The density computed at the mean-field (not shown) and many-body levels depict identical features for all $\omega_r$. All quantities shown are dimensionless.}\label{density_three}
 \end{figure}

We computed the ground-state energy $E/N$, and found it to display similar pattern as that for the elongated trap. 
That is for slow rotation, E/N remains almost constant and then E/N drops gradually with increase in $\omega_r$.
In addition, we found that the mean-field and many-body energies are practically identical for all $\omega_r$. 
The results are shown in the supplemental material, see Figure S1(b).

\begin{figure}[htbp]
  \centering
  \includegraphics[angle=270,width=0.45\textwidth]{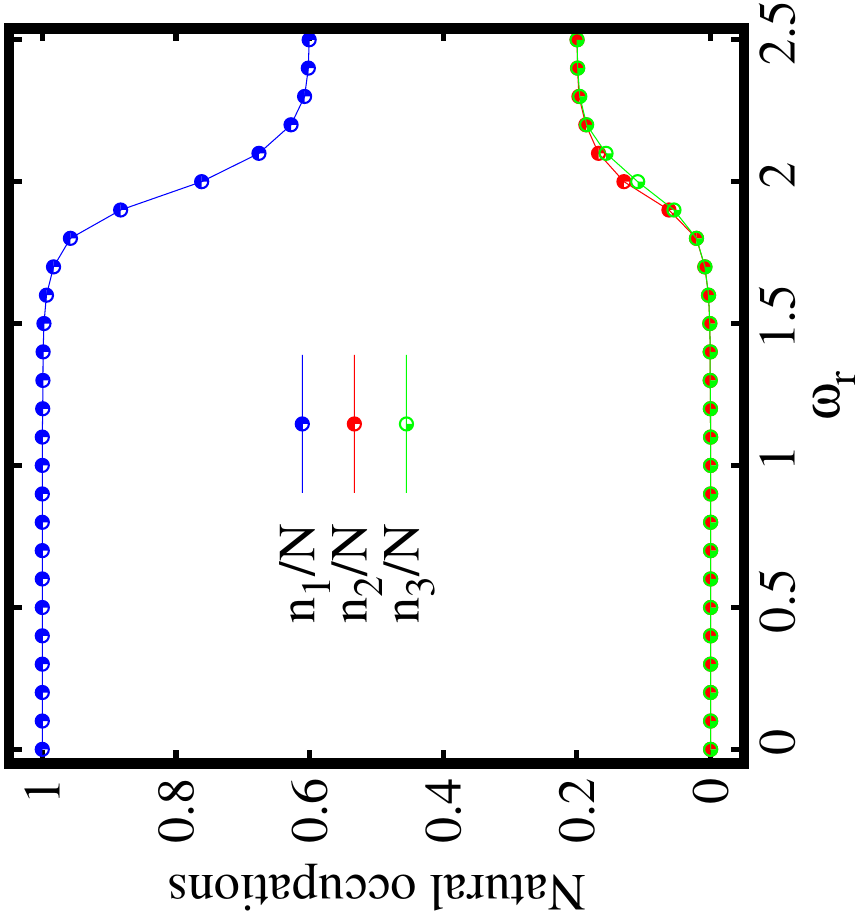}
 \caption{Pathway from condensation to fragmentation in the rotating three-fold symmetric trap.
 three-fold fragmentation is observed with increase in rotation. The three leading natural occupations $n_1/N$, $n_2/N$ and $n_3/N$, are shown as a function of the rotation frequency $\omega_r$. $M=3$ self-consistent orbitals are used. All quantities shown are dimensionless.
 }\label{occupation_three}
 \end{figure}
 Figure (\ref{density_three}) displays the ground-state densities per particle of a rotating condensate for three different rotation frequencies $\omega_r$, confined in the three-fold symmetric trap [Equation \eqref{eq:potentials_three}] computed at the many-body level. In the absence of rotation, the one-body density depicts a single cloud where all the bosons accumulate in the center of the trap. However, faster rotations lead to breakup of the density profile into three clouds. We have also computed the ground state density at the mean-field level and it shows essentially identical features as the many-body level density in real space.
 
To further understand the breaking of the ground-state density, we discuss the behaviour of the natural occupations $\frac{n_j}{N}$ as a function of the rotation frequency $\omega_r$ as shown in Figure \ref{occupation_three}. 
It is evident from Figure (\ref{occupation_three}) that the system remains fully condensed, i.e., $\frac{n_1}{N}\sim1, \frac{n_2}{N}\sim \frac{n_3}{N}\sim10^{-6}$  with the inclusion of rotation till about $\omega_r=1.1$. 
From about $\omega_r=1.2$ onwards, the condensate starts to deplete with gradual decrease in the population of the first natural orbital followed by corresponding increase in the populations of the second and third natural orbitals, $\frac{n_2}{N}\sim \frac{n_3}{N}\sim 10^{-4}$ [see Figure S2(b) in the supplemental material]. 
For a faster rotation, say $\omega_r=2$, the system transits to a fragmented state with decrease in the population of the first natural orbital followed by macroscopic population of the second and third natural orbitals. 
Further increase in the rotation to $\omega_r=2.5$ leads to a three-fold fragmented state having the natural occupations $\frac{n_1}{N}\sim60\%$ and $\frac{n_2}{N}\sim \frac{n_3}{N}\sim 20\%$ of the first, second, and third natural orbitals, respectively. 
It is observed that to achieve a fully three-fold fragmentation in this three-fold symmetric trap, we would need a much faster rotation in comparison with the elongated trap.
Therefore, in case of a weakly interacting bosons confined in the three-fold symmetric trap given by Equation \eqref{eq:potentials_three}, switching on rotation leads to a transition from a fully condensed state to three-fold fragmentation. 

 \begin{figure}[!t]
  \centering
  \includegraphics[angle=270,width=0.45\textwidth]{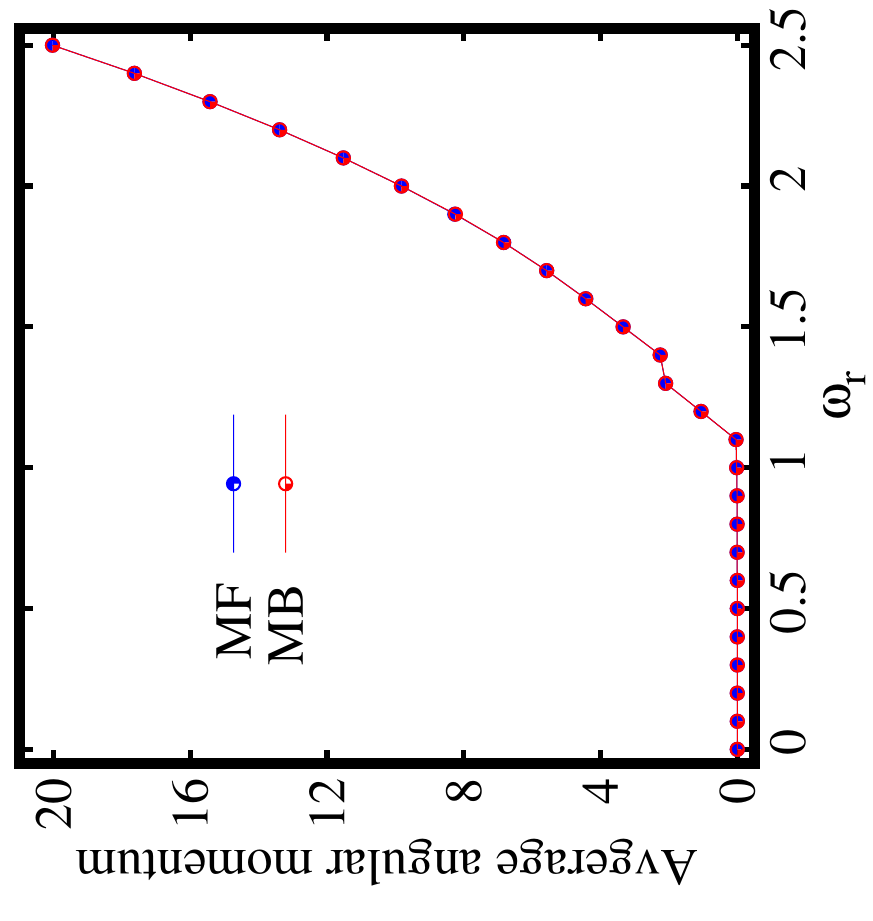}
 \caption{Expectation value of angular momentum operator $\langle \hat{L}_Z \rangle/N$, computed at the mean-field (MF) and many-body (MB) levels [with $M=1$ and $M=3$ self-consistent orbitals, respectively] as a function of the rotation frequency $\omega_r$ for the three-fold symmetric trap. All quantities shown are dimensionless.}\label{mom_three}
 \end{figure}

Let us move to the behaviour of average angular-momentum per particle $\langle \hat{L}_Z \rangle/N$, as a function of the rotation frequency $\omega_r$ for the three-fold symmetric trap computed both at the mean-field and many-body levels.
 It is observed from Figure (\ref{mom_three}) that the mean-field and many-body angular momenta exactly match each other for all $\omega_r$. 
 The angular momentum remains minimum till the rotation frequency of about $\omega_r=1.1$. For about $\omega_r=1.2$, the rotation produces a state where significant value of angular momentum generates with $\langle \hat{L}_Z\rangle/N> 1$ at which the breakup of the density is observed. 
 The angular momentum gradually increases with further increase in the rotation of the condensate. Even for strong rotation, the angular momentum computed at the mean-field and many-body levels coincide each other. 
 Therefore, we can conclude that, at least for the ground state, the angular momentum and its variance [see Figure S6(b) in the supplemental material] are not good indicators for many-body effects.
 Nonetheless, we stress that the acquisition of angular momentum in the condensate follows the breakup of the density, the emergence of depletion, and the eventual fragmentation in the three-fold symmetric trap.
 \begin{figure}[htbp]
  \centering
  \includegraphics[angle=270,width=0.45\textwidth]{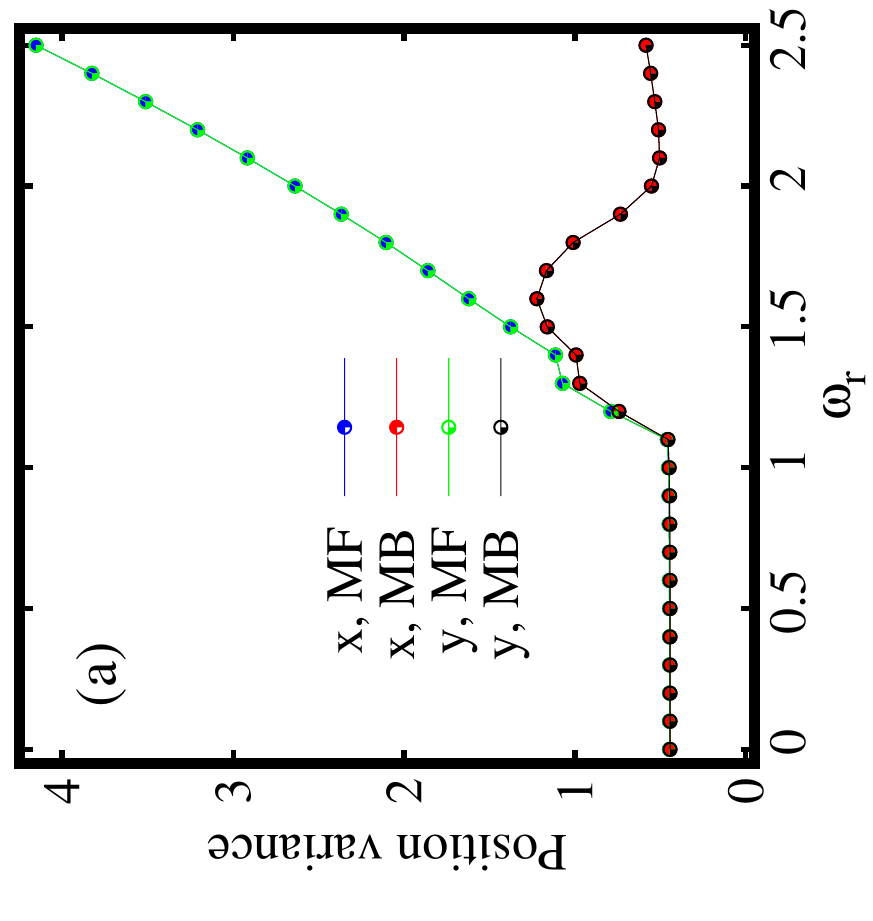}
 \includegraphics[angle=270,width=0.45\textwidth]{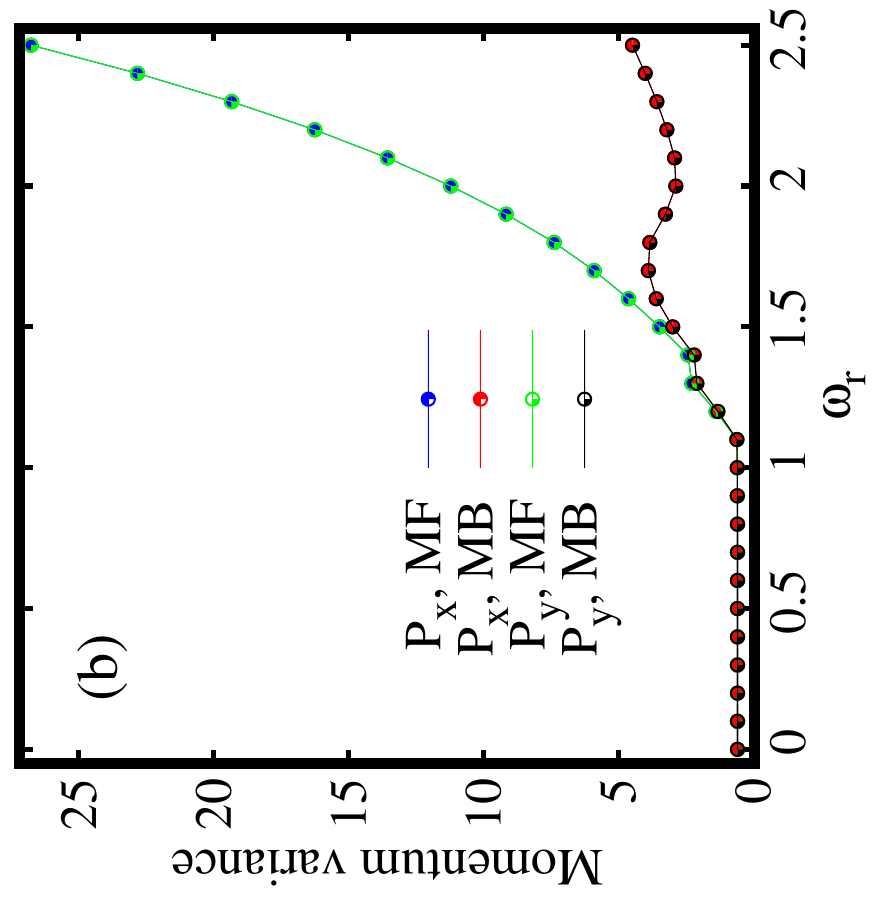}
 \caption{Dependence of the the many-particle position and momentum variances on the rotation in the three-fold symmetric trap. Shown are (a) $\frac{1}{N}\Delta^2_{\hat{X}}$ and   $\frac{1}{N}\Delta^2_{\hat{Y}}$ and (b) $\frac{1}{N}\Delta^2_{\hat{P}_X}$ and $\frac{1}{N}\Delta^2_{\hat{P}_Y}$ as a function of $\omega_r$ at the many-body level (MB) [$M=3$ self-consistent orbitals] and at the mean-field level (MF) [$M=1$ self-consistent orbital]
  All quantities shown are dimensionless.}\label{variance_three}
 \end{figure}

Now we move to the discussion of the impact of rotation on the behaviour of the many-particle variances of the position and momentum operators along the $x$- and $y$-directions, to further characterize the many-body properties of the rotating condensate undergoing breakup.

 Figure \ref{variance_three}(a) displays the behaviour of the many-particle position variance per particle $\frac{1}{N}\Delta^2_{\hat{X},\hat{Y}}$ as a function of the rotation frequency $\omega_r$ in the three-fold symmetric trap computed at the mean-field and many-body levels. 
 It is observed that the mean-field and many-body $\frac{1}{N}\Delta^2_{\hat{X},\hat{Y}}$ coincide till about $\omega_r=1.1$. 
 For faster rotations, from about $\omega_r=1.2$ onwards the mean-field and many-body position variances deviate. 
 The mean-field $\frac{1}{N}\Delta^2_{\hat{X},\hat{Y}}$ increase monotonously depicting the spreading and finally the breakup of the density. 
 However, the many-body $\frac{1}{N}\Delta^2_{\hat{X},\hat{Y}}$ first increase and then, after reaching a maximal value starts decreasing with further increase in $\omega_r$, because of the depletion and eventual fragmentation. 
 One of the important features of the position variance is that the variance along the $x$- and $y$-direction exactly coincide each other for both the mean-field and many-body regimes, thereby indicating the fact that the three-fold rotational symmetry of the condensate is preserved even for fast rotations.
 Clearly, there is no anisotropy of the variance for all $\omega_r$ in this trap.
 
 Finally, Figure \ref{variance_three}(b) shows the behaviour of the momentum variance per particle $\frac{1}{N}\Delta^2_{\hat{P}_X,\hat{P}_Y}$, computed at the mean-field and many-body levels as a function of $\omega_r$. 
 The momentum variances remain small even with increase in rotation till about $\omega_r=1.1$. 
 From about $\omega_r=1.2$ onwards, the momentum variances computed both at the mean-field and many-body levels slowly start to increase. 
 Here, the mean-field momentum variance only increases monotonously. 
The many-body momentum variance increases, decreases, and again increases but remain much smaller than the mean-field momentum variance, indicates the presence of depletion and fragmentation.
 Similar to the position variance, the momentum variance along the $x$- and $y$-directions exactly coincide each other both at the mean-field and many-body levels, due to the rotational symmetry.
 
\subsubsection{Bosons in a four-fold symmetric trap under rotation}
\begin{figure}[htbp]
  \centering
   \includegraphics[angle=270,width=1.0\textwidth]{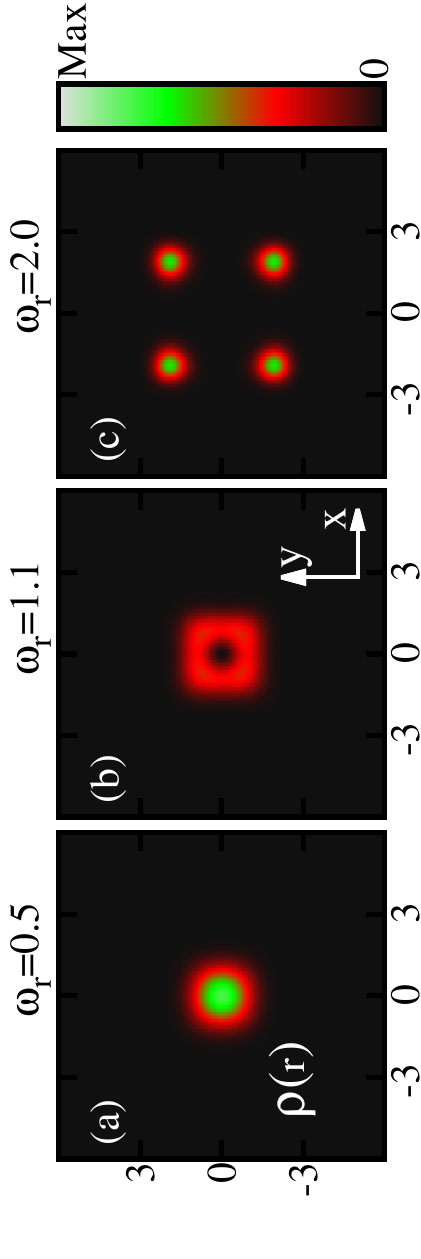}
  \caption{The one-body densities per particle are shown for three different rotation frequencies $\omega_r$ in the four-fold symmetric trap at the many-body level. M=4 self-consistent orbitals are used. The density computed at the mean-field (not shown) and many-body levels display identical features at all $\omega_r$. All quantities shown are dimensionless.}\label{density_four}
 \end{figure}
 Now we move to a more complicated system, a four-fold symmetric trap, to show the stability of the ground-state properties found above for the three-fold symmetric trap. The potential of the four-fold symmetric trap is given by
 \begin{eqnarray}
 V(\mathbf{r}) = \frac{1}{4}(x^4+y^4).   
\label{eq:potentials_four}
\end{eqnarray}
Here, we consider $M = 4$ self-consistent orbitals to obtain the ground-state properties of bosons under rotation in this trap. We also compute the results with $M=8$ self-consistent orbitals to verify the numerical convergence, see the supplemental material. The range of rotation frequencies is taken to be $\omega_r=[0,2.0]$ for the following study.

 \begin{figure}[htbp]
  \centering
  \includegraphics[angle=270,width=0.45\textwidth]{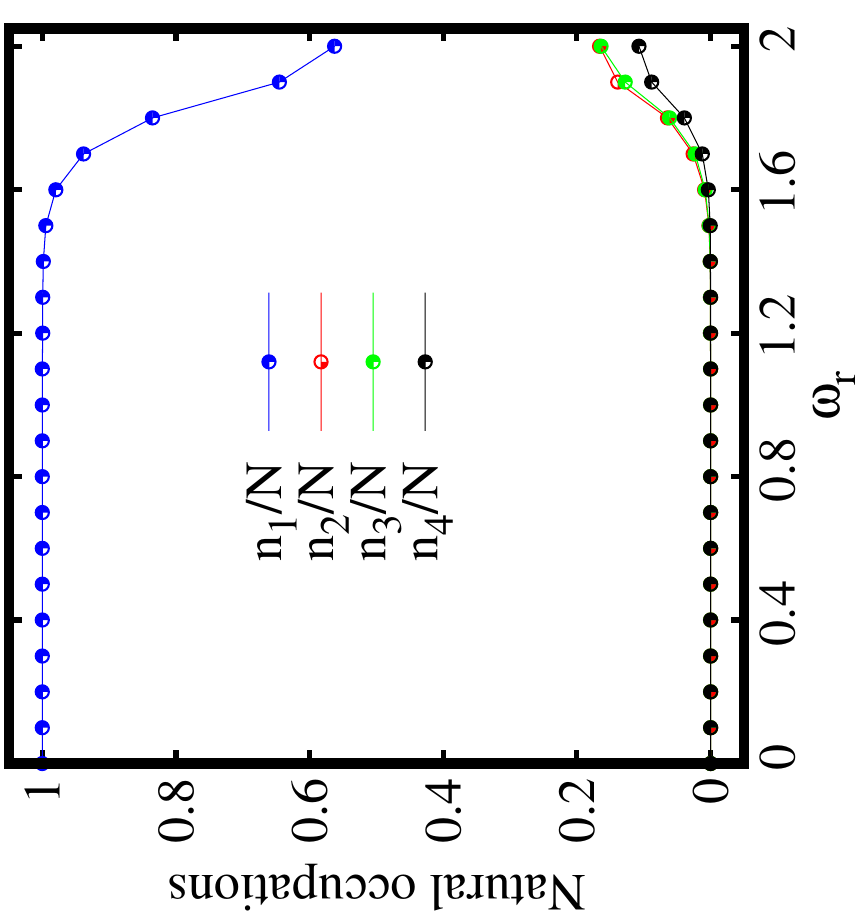}
 \caption{Rotation leads to four-fold fragmentation in the four-fold symmetric trap.
 $M=4$ self-consistent orbitals are used.
 The variation of four natural occupations $n_1/N$, $n_2/N$, $n_3/N$ and $n_4/N$ are shown as a function of rotation frequency $\omega_r$. All quantities computed are dimensionless.}\label{occupation_four}
 \end{figure}
We computed the ground state energy $E/N$, and found it to display similar pattern as that for the elongated and three-fold symmetric traps. 
That is for slow rotation, E/N remains almost constant and then E/N drops gradually with increase in $\omega_r$.
In addition, we found that the mean-field and many-body energies are practically identical for all $\omega_r$. 
The results are shown in the supplemental material, see Figure S1(c). 
 
Figure \ref{density_four} shows the behaviour of the ground-state densities per particle of a rotating BEC confined in the four-fold symmetric trap for three different rotation frequencies $\omega_r$ at the many-body level. Similar to the elongated and three-fold symmetric traps, for slow rotation the density displays a single cloud. With increasing rotation, a deep in the density emerges [Figure \ref{density_four}(b)] and finally, faster rotations lead to splitting of the density into four sub-clouds as evident from Figure \ref{density_four}(c).

For a deeper understanding of the many-boson density profile, the behaviour of the natural
occupations $\frac{n_j}{N}$ as a function of the rotation frequency $\omega_r$ is analyzed in Figure (\ref{occupation_four}). 
\begin{figure}[htbp]
  \centering
  \includegraphics[angle=270,width=0.45\textwidth]{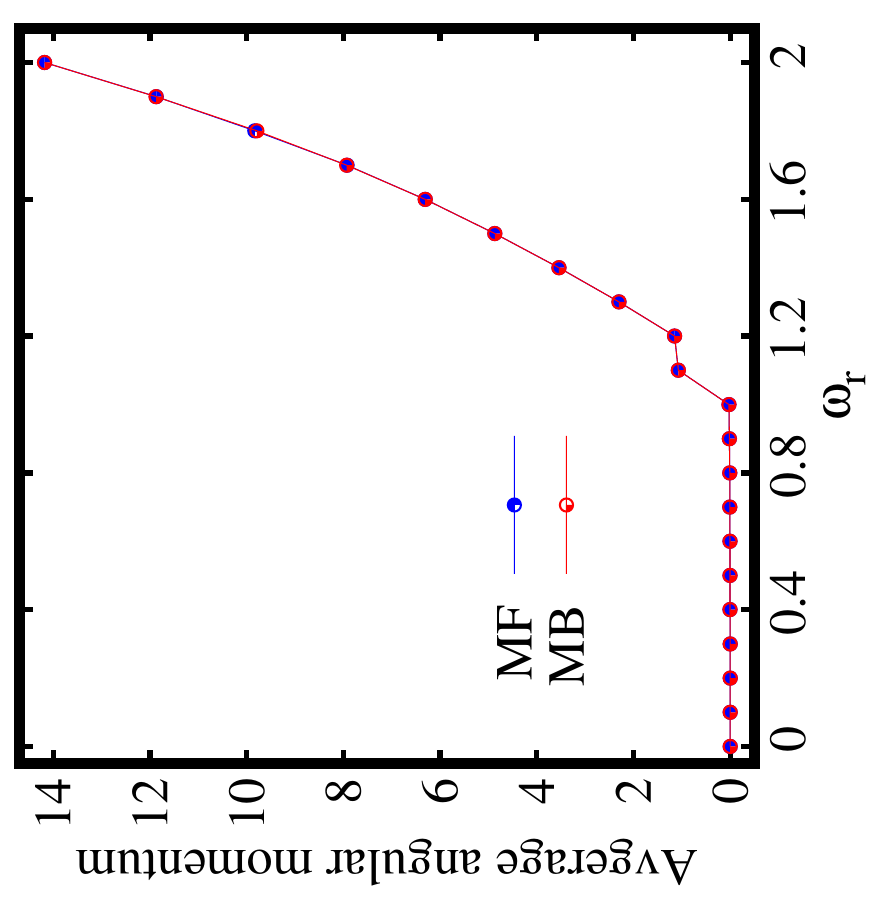}
 \caption{Expectation value of angular momentum operator $\langle \hat{L}_Z \rangle/N$, computed at the mean-field (MF) and many-body (MB) levels [with $M=1$ and $M=4$ self-consistent orbitals, respectively] as a function of the rotation frequency $\omega_r$ for the four-fold symmetric trap. All quantities shown are dimensionless.}\label{ang_mom_four}
 \end{figure}
 It is found that the system preserves the fully condensed state, with $\frac{n_1}{N}\sim1, \frac{n_2}{N}\sim\frac{n_3}{N}\sim \frac{n_4}{N}\le 10^{-6}$, till a rotation frequency of about  $\omega_r=1.0$. Further increase in $\omega_r$ leads to slow depletion of the condensate. For a faster rotation, $\omega_r=2.0$, four-fold fragmentation of the condensate with finite population of all the four natural orbitals is observed. 

 Let us discuss the behaviour of the average angular momentum per particle as a function of $\omega_r$, see Figure \ref{ang_mom_four}. 
The angular momentum remains minimal for $\omega_r=1$. From about $\omega_r=1.1$, the rotation produces a state where significant value of angular momentum generates with $\langle \hat{L}_Z\rangle/N> 1$ at which the breakup of the density is observed. 
 The angular momentum gradually increases with further increase in the rotation. Even for strong rotation, the angular momenta computed at the mean-field and many-body levels coincide each other. Finally, it can be concluded for the four-fold trap as well that, at least for the ground state, the angular momentum and its variance [see Figure S6(C) in the supplemental material] do not precisely signifies many-body effects.
 However, we can conclude that accumulation of angular momentum in the condensate, the breakup of the density, emergence of depletion, and eventual fragmentation in the four-fold symmetric trap are in sync similar to the elongated and three-fold symmetric traps.

\begin{figure}[htbp]
  \centering
  \includegraphics[angle=270,width=0.45\textwidth]{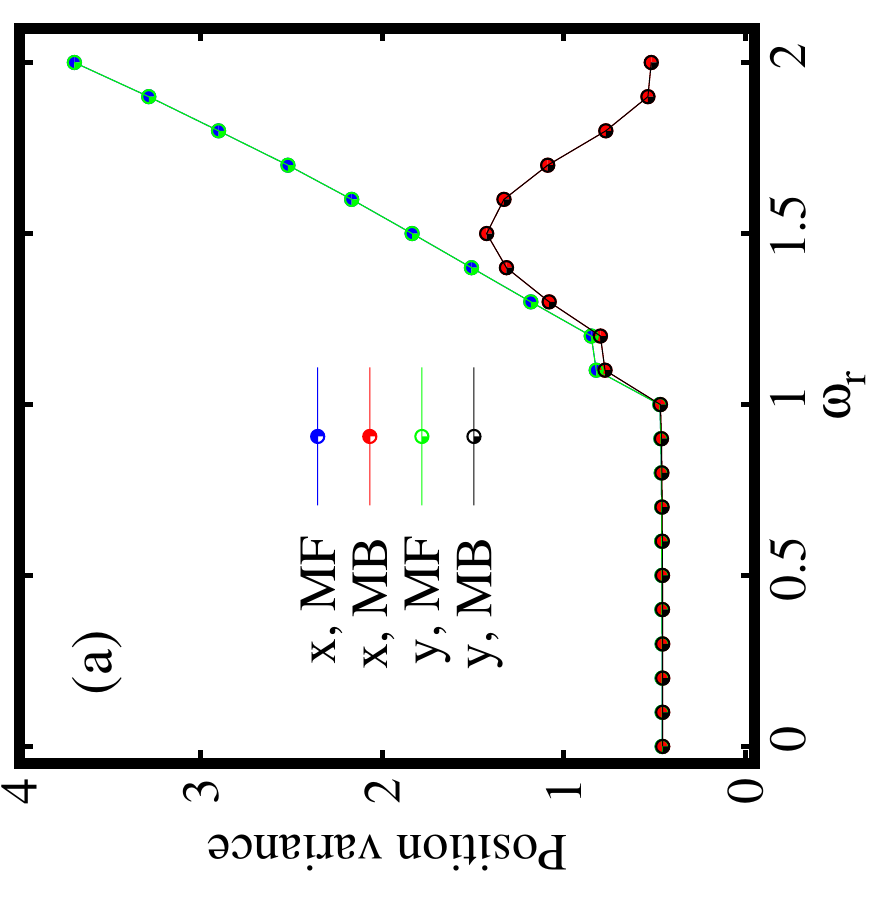}
  \includegraphics[angle=270,width=0.45\textwidth]{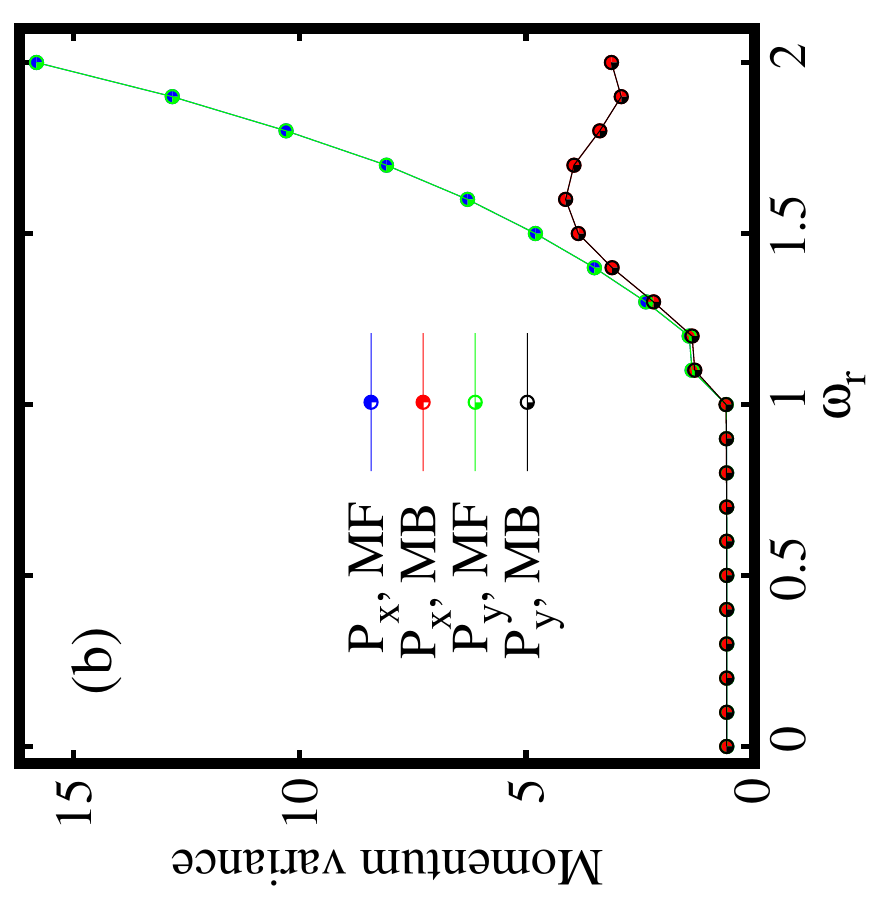}
 \caption{Dependence of the the many-particle position and momentum variances on the rotation in the four-fold symmetric trap.
 (a) depicts the position variances $\frac{1}{N}\Delta^2_{\hat{X}}$, $\frac{1}{N}\Delta^2_{\hat{Y}}$ and (b) displays momentum variances $\frac{1}{N}\Delta^2_{\hat{P}_X}$, $\frac{1}{N}\Delta^2_{\hat{P}_Y}$ at the many-body level (MB) [$M=4$ self-consistent orbitals] and at the mean-field level (MF) [$M=1$ self-consistent orbital]. All quantities shown are dimensionless.}\label{variance_four}
 \end{figure}
Figures \ref{variance_four} displays the behaviour of the position and momentum variances computed at the mean-field and many-body levels as a function of rotation frequency $\omega_r$ for four-fold symmetric trap along the $x$- and $y$-directions. 
The mean-field and many-body position variances $\frac{1}{N}\Delta^2_{\hat{X},\hat{Y}}$ coincide till about $\omega_r=1$. 
For faster rotations, from about $\omega_r=1.1$ onward, the mean-field position variance deviates from the many-body position variance and the former is always larger than the latter. 
Further, the position variances along the $x$- and $y$-directions exactly coincide with each other both at the mean-field and many-body levels similar to the three-fold symmetric trap, see Figure \ref{variance_three}(a).
Side by side, the behaviour of the momentum variance $\frac{1}{N}\Delta^2_{\hat{P}_X,\hat{P}_Y}$ displays similar feature as that of the position variance, both at the mean-field and many-body levels. 
In particular, mean-field is larger than many-body from about $\omega_r=1.5$.

Finally, we can conclude that the depletion, angular momentum, and the position and momentum variances follow hand in hand both in the three-fold and four-fold symmetric traps. 
The rotating interacting bosons acquire unique many-body properties while undergoing breakup in space.

\section{Concluding remarks}
In the present work, we have studied the impact of rotation on the ground state of weakly interacting bosonic atoms confined in two-dimensional anharmonic potentials, first in, an elongated trap and, subsequently, in three-fold and four-fold symmetric traps.
Here, the multiconfigurational time-dependent Hartree method for bosons, which is particularly suitable to describe many-body properties, is employed to investigate the ground-state energy, density, the depletion and fragmentation, angular momentum, and finally, many-particle variances as a function of the rotation frequency to characterize the correlations present in the system. 

In the elongated trap, it is observed that the ground-state density breaks up into two clouds with rotation. The splitting of the density is followed by the emergence of the two-fold fragmentation. Interestingly, the ground state exhibits opposite anistoropy both for the many-particle position and momentum variances when computed at the many-body and mean-field levels.
The rotation squeezes the position variance in the elongated direction and the momentum variance in the narrow direction, thereby producing unique correlations.
Finally, a synchronized pattern among the density breakup, eventual fragmentation, acquisition of angular momentum in the condensate, and many-particle variances is observed.
When the angular momentum sets in the breakup of density and the many-particle position and momentum variances start to increase.

For the three-fold and four-fold symmetric traps, the ground-state density eventually splits into three and four clouds, respectively, with the inclusion of the rotation. Side-by-side, the rotation leads to transition from condensed to three-fold and four-fold fragmented condensates, respectively, at the many-body level of theory. We find that the depletion, the accumulation of angular momentum, and finally, the increase in the variances of position and momentum follow hand in hand.

The rotating frame in our work can be viewed as a specific case of synthetic gauge fields.
In future, continuation of this investigation includes the extension of the study in various synthetic gauge fields. It would be interesting to explore the ground-state breakup, eventual condensation or fragmentation, and finally, various correlations of the condensate in presence of synthetic gauge fields. It would be fascinating to discover many-body features beyond the capacity of a ``simple" rotation.
\nocite{*}
\section{Acknowledgements}
This work is supported by the Israel Science Foundation (ISF) grant no. 1516/19 and by the Austrian Science Foundation (FWF) under grant P-32033-N32.
 SD acknowledges Anal Bhowmik for helpful discussions. 
Computation
time at the High-Performance Computing Center Stuttgart (HLRS) is gratefully
acknowledged.
\bibliography{manuscript} 
\newpage
\section*{Supplemental material to fragmentation and correlations in a rotating Bose-Einstein condensate undergoing breakup}
This Supplemental material provides further many-body analysis that supports our main
results. Sec. S1 benchmarks the multiconfigurational time-dependent Hartree for bosons (MCTDHB) method \cite{lode2012numerically,two,lode:20} 
in a harmonic
interaction model (HIM) under rotation that gives an exactly solvable many-body model. This method is employed for the numerical computations of the main results.
Sec. S2 gives the convergence checks of the ground state energy per particle of our main results with respect to the number of self-consistent orbitals in different anharmonic traps, namely, elongated, three-fold symmetric, and four-fold symmetric traps.
Sec. S3 illustrates the convergence of natural occupations in these confining traps with respect to the number of orbitals. Sec. S4 presents the many-particle variances discussed in the main text, along with
their convergences with respect to the number of  self-consistent orbitals. This section is further divided into two subsections. The first subsection deals with convergence of the position and momentum variances. Convergence with respect to the number of self-consistent orbitals of the expectation value of the angular momenta  and angular momentum variance is elaborated in the second subsection that supports our main results.
\section{Benchmarking of MCTDHB in the rotating frame: An exactly Solvable many-body model} \label{sec:method}
 In this section, we present a numerical benchmark of the MCTDHB method in the rotating frame in two dimensions 2D. The many-body Hamiltonian in the rotating frame can be written as
 \begin{equation}
 \hat{H}(\mathbf{r_1}, \mathbf{r_2},...,\mathbf{r_N})=\sum_{j=1}^N\hat{h}(\mathbf{r}_i)+\sum_{j<k}^N\hat{W}(\mathbf{r}_j,\mathbf{r}_k)-\omega_r \hat{L}_Z,
\label{eq:hamiltonian}
\end{equation}
where the single-particle Hamiltonian is $\hat{h}(\mathbf{r})=-\frac{1}{2}\frac{\partial^2}{\partial \mathbf{r}^2}+\hat{V}(\mathbf{r})$. We consider a harmonic interaction model (HIM) in 2D. In this HIM, the 
 confining trapping potential $\hat{V}(\mathbf{r})$ is a translated 2D harmonic oscillator of the following form
\begin{equation}
 \hat{V}(\mathbf{r})=\frac{1}{2}m\omega^2[(x-L)^2+y^2],
\label{eq:potentials_him}
\end{equation}
and the two-body
interaction potential $\hat{W}(\mathbf{r}_j,\mathbf{r}_k)$ is also harmonic
\begin{eqnarray}
\hat{W}(\mathbf{r}_j,\mathbf{r}_k)= \lambda_0(\mathbf{r}_j-\mathbf{r}_k)^2,
\end{eqnarray}
 where the two-body interaction strength $\lambda_0$ defines the mean-field interaction parameter as $\Lambda=\lambda_0(N-1)$,
$\omega_r$ is the rotation frequency, and $\hat{L}_Z=\sum_{j=1}^N\hat{l}_{z_{j}}=\sum^N_{j=1}\left( \hat{x}_j \hat{p}_{y_j} - \hat{y}_j\hat{p}_{x_j} \right)$ is the many-body angular momentum operator.

Now, the ground state solution of the HIM in the rotating frame is obtained by solving the many-boson Schr\"{o}dinger equation given as
\begin{equation}
\hat{H}(\mathbf{r})\Psi(\mathbf{r})=E\Psi(\mathbf{r}).   
\end{equation}
The $N$-boson wave function of the ground state in the rotating frame is obtained as
\begin{equation}
\Psi(\mathbf{r}_1,\mathbf{r}_2,...,\mathbf{r}_N)=\bigg( \frac{m\Omega}{\pi} \bigg)^{\frac{N-1}{4}}\bigg(\frac{m\omega}{\pi} \bigg)^{\frac{1}{4}} e^{+i \beta \sum_{j=1}^{N} y_j}e^{-\frac{1}{2}m\omega\sum_{j=1}^{N}(x_j-x_0)^2}e^{-\frac{1}{2}m\omega \sum_{j=1}^{N}y_j^2},
\label{wave}
\end{equation}
where the dressed frequency is
\begin{equation}
 \Omega=\sqrt{\omega^2+\frac{2\lambda_0 N}{m}}, \nonumber
\end{equation}
and the translation and momentum parameters are
\begin{equation}
x_0 = \frac{1}{1-(\frac{\omega_r}{\omega})^2}L    
\end{equation}
and
\begin{equation}
\beta = m\omega_r x_0= m\omega_r\frac{1}{1-(\frac{\omega_r}{\omega})^2}L.    
\end{equation}
Thus, we can evaluate the ground state properties, such as, the energy, densities,
and reduced density matrices from the many-boson wavefunction $\Psi$ in Equation \eqref{wave}. 

As we know, the ground state energy per particle in the HIM without rotation is given as \cite{six}
\begin{equation}
 \frac{E(\omega_r=0)}{N}=\bigg[(N-1)\Omega+\omega\bigg]=\bigg[(N-1)\sqrt{\omega^2+\frac{2\lambda_0 N}{m}}+\omega\bigg].
\end{equation}
Finally, in the rotating frame the energy per particle is obtained as
\begin{equation}
 \frac{E(\omega_r)}{N}=\frac{E(\omega_r=0)}{N}-\frac{1}{2}m\omega^2  \frac{(\frac{\omega_r}{\omega})^2}{1-(\frac{\omega_r}{\omega})^2}L^2.
\end{equation}

Now we are going to evaluate the expectation values and variances of many-particle observables, explicitly the position, momentum, and angular momentum operators.
The expectation values of the many-particle position operators are given as
\begin{equation}
\frac{\langle\hat{X}\rangle}{N}=x_0= \frac{1}{1-(\frac{\omega_r}{\omega})^2}L,~~~~~~~ \frac{\langle\hat{Y}\rangle}{N}=0,
\end{equation}
and the expectation values of the many-particle momentum operators are given as
\begin{equation}
\frac{\langle\hat{P}_X\rangle}{N}=0,~~~~~~~ \frac{\langle\hat{P}_Y\rangle}{N}=\beta = m\omega_r x_0= m\omega_r\frac{1}{1-(\frac{\omega_r}{\omega})^2}L .
\end{equation}
Similarly, the expectation value of the many-particle angular momentum operator reads
\begin{equation}
 \frac{\langle\hat{L}_Z\rangle}{N}=x_0\beta= m\omega_r \Bigg[ \frac{L}{1-(\frac{\omega_r}{\omega})^2}\Bigg]^2. 
\end{equation}
The many-particle position variances per particle are obtained as,
\begin{equation}
\frac{1}{N}\Delta^2_{\hat{X}}=\frac{1}{N}\Delta^2_{\hat{Y}}=\frac{1}{2m\omega},
\end{equation}
and the many-particle momentum variances per particle are the following
\begin{equation}
\frac{1}{N}\Delta^2_{\hat{P}_X}=\frac{1}{N}\Delta^2_{\hat{P}_Y}=\frac{m\omega}{2}.
\end{equation}
It is evident that the many-particle variances of the position and momentum operators in the rotating frame boil down to those for the non-rotating case. 

However, the many-particle angular momentum variance per particle is affected by the rotation and takes on the following form
\begin{eqnarray}
\frac{1}{N}\Delta^2_{\hat{L}_Z}=x_0^2\frac{1}{N}\Delta^2_{\hat{P}_Y}+\beta^2\frac{1}{N}\Delta^2_{\hat{X}}
=\frac{1}{2}m\omega\frac{\bigg [1+(\frac{\omega_r}{\omega})^2\bigg]}{\bigg[1-(\frac{\omega_r}{\omega})^2\bigg]^2} L^2.
\end{eqnarray}

With a solvable many-boson model in the rotating frame, we can now proceed and benchmark MCTDHB in the rotating frame.
In the following simulations, we work in the units $\hbar=m=\omega=1$.
To compute the ground state of MCTDHB in the rotating frame,  we consider $N=10$ weakly interacting bosons with repulsive interaction parameter $\Lambda=-0.1$. The box extensions of the confining potential are chosen to be $ \bf {= [-12,12)\times[-12,12)}$ with $\bf{128\times 128}$ DVR functions to represent each of the orbitals. We translated the confining potential by $L=2$ in Equation \eqref{eq:potentials_him}. 

In Table \ref{tab:1}, we show the numerical convergence of the energy per particle $\frac{E}{N}$. 
Exact analytical versus numerical MCTDHB ground state energies are obtained for three different rotation frequencies $\omega_r$. The numerical convergence of the many-body energy per particle is achieved by using $M=6$ self-consistent orbitals. Thus, we conclude that for all $\omega_r$, the ground state energy of the system is converged.
\begin{table}[htbp]
\begin{center}
\begin{tabular}{|c|c|c|}\hline
$\omega_r$ &  ${\frac{E}{N}}_{numerical}$ &$ {\frac{E}{N}}_{analytical}$
\\\hline            
$0$  &     0.89372 &0.89372 \\\hline
$0.5$ &    0.22705& 0.22705\\\hline
$0.8$  &  -2.66183&-2.66183\\\hline   
\end{tabular}
\end{center}
\caption{\label{tab:1} Benchmarking of the ground state many-body energy E/N with respect to the rotation frequency $\omega_r$. $M=6$ self-consistent orbitals are used. The ground state energies shown are in dimensionless units.}
\end{table}
\begin{table}[htbp]
\begin{center}
\begin{tabular}{|c|c|c|c|c|}\hline
 $\omega_r$ & ${\frac{1}{N} \langle \hat{X} \rangle}_{numerical}$ &  ${ \frac{1}{N}\langle \hat{X} \rangle}_{analytical}$& ${ \frac{1}{N}\langle \hat{Y} \rangle}_{numerical}$& ${ \frac{1}{N}\langle \hat{Y} \rangle}_{analytical}$\\\hline
 0 &   2.00000 &2.00000 &  0.00000 & 0.00000\\\hline
 0.5 & 2.66666 &2.66666 & 0.00000  & 0.00000 \\\hline
  0.8 &5.55555& 5.55555 &  0.00000  & 0.00000 \\\hline
      
$\omega_r$ & ${ \frac{1}{N} \Delta^2_{\hat{X}}}_{numerical}$ & ${ \frac{1}{N} \Delta^2_{\hat{X}}}_{analytical}$& ${ \frac{1}{N} \Delta^2_{\hat{Y}}}_{numerical}$ & ${ \frac{1}{N} \Delta^2_{\hat{Y}}}_{analytical}$\\\hline
 0 &0.50000 & 0.50000 &   0.50000  &0.50000 \\\hline
 0.5& 0.50000 & 0.50000 & 0.50000  & 0.50000\\\hline
  0.8 & 0.50000 & 0.50000 & 0.50000  & 0.50000\\\hline
\end{tabular}          
\end{center}
\caption{\label{tab:2} Benchmarking of the expectation values and variances of the many-particle position operators along the $x$- and $y$- directions with respect to the rotation 
frequency $\omega_r$ computed with $M=6$ self-consistent orbitals. All the quantities are shown in dimensionless units.}
\end{table}

Table \ref{tab:2} shows the comparison of the exact analytical and numerical MCTDHB results of the many-particle position variances for various rotation frequencies $\omega_r$. The numerical convergence of MCTDHB results computed with $M=6$ self-consistent orbitals are clearly evident from the table.
\begin{table}[htbp]
\begin{center}
\begin{tabular}{|c|c|c|c|c|}\hline
 $\omega_r$ & ${ \frac{1}{N} \langle \hat{P}_X \rangle}_{numerical}$ &  ${ \frac{1}{N} \langle \hat{P}_X \rangle}_{analytical}$&${ \frac{1}{N} \langle \hat{P}_Y \rangle}_{numerical}$& ${\frac{1}{N} \langle \hat{P}_Y \rangle}_{analytical}$\\\hline
 0 &  0.00000 & 0.00000 &0.00000& 0.00000 \\\hline
 0.5 &0.00000 & 0.00000 &1.33333& 1.33333\\\hline
  0.8&0.00000 & 0.00000 &4.44444& 4.44444\\\hline
$\omega_r$ & ${\frac{1}{N} \Delta^2_{\hat{P}_X}}_{numerical}$& ${ \frac{1}{N} \Delta^2_{\hat{P}_X}}_{analytical}$ & ${\frac{1}{N} \Delta^2_{\hat{P}_Y}}_{numerical}$ & ${\frac{1}{N} \Delta^2_{\hat{P}_Y}} _{analytical}$\\\hline
0   &0.50000 &0.50000 &0.50000& 0.50000\\\hline
0.5 &0.50000& 0.50000& 0.50000& 0.50000\\\hline
0.8 &0.50000& 0.50000 &0.50000& 0.50000\\\hline
 \end{tabular}
\end{center}
\caption{\label{tab:3} Benchmarking of the expectation values and variances of the many-particle momentum operators along the $x$- and $y$- directions with respect to the rotation 
frequency $\omega_r$ obtained with $M=6$ self-consistent orbitals. All the quantities are shown in dimensionless units.}
\end{table}
\begin{table}[htbp]
\begin{center}
\begin{tabular}{|c|c|c|c|c|}\hline
$\omega_r$ & ${ \frac{1}{N} \langle \hat{L}_Z \rangle}_{numerical}$ & ${ \frac{1}{N} \langle \hat{L}_Z \rangle}_{analytical}$ &  ${ \frac{1}{N} \Delta^2_{\hat{L}_Z}}_{numerical}$ & ${ \frac{1}{N} \Delta^2_{\hat{L}_Z}}_{analytical}$ \\
\hline                    
0 & 0.00000  & 0.00000  & 2.00000&2.00000\\\hline
 0.5 &3.55555& 3.55555 &  4.44444 &4.44444\\\hline
 0.8 &24.6913& 24.6913 &25.3086&25.3086 \\\hline 
\end{tabular}
\end{center}
\caption{\label{tab:4} Benchmarking of the expectation values and variances of the many-particle angular momentum operator with respect to the rotation 
frequency $\omega_r$ computed with $M=6$ self-consistent orbitals. All the quantities are shown in dimensionless units.}
\end{table}

Table \ref{tab:3} and \ref{tab:4} correspond to benchmarking of the many-particle momentum and angular momentum variances for various rotation frequencies computed with $M=6$ self-consistent orbitals. As the many-particle variances of the momentum and angular momentum are very sensitive quantities, it is generally more difficult to achieve numerical convergence of the analytical and numerical results. However, from these tables, it is evident that the MCTDHB results are converged with exact analytical results.

\section{Convergence of the ground state energy}\label{sec:convergence}
In this section, we investigate and report the convergence of the many-particle ground state energy $\frac{E}{N}$
for the three confining potentials,
\begin{figure}[!t]
 \centering
 \includegraphics[angle=270,width=0.48\textwidth]{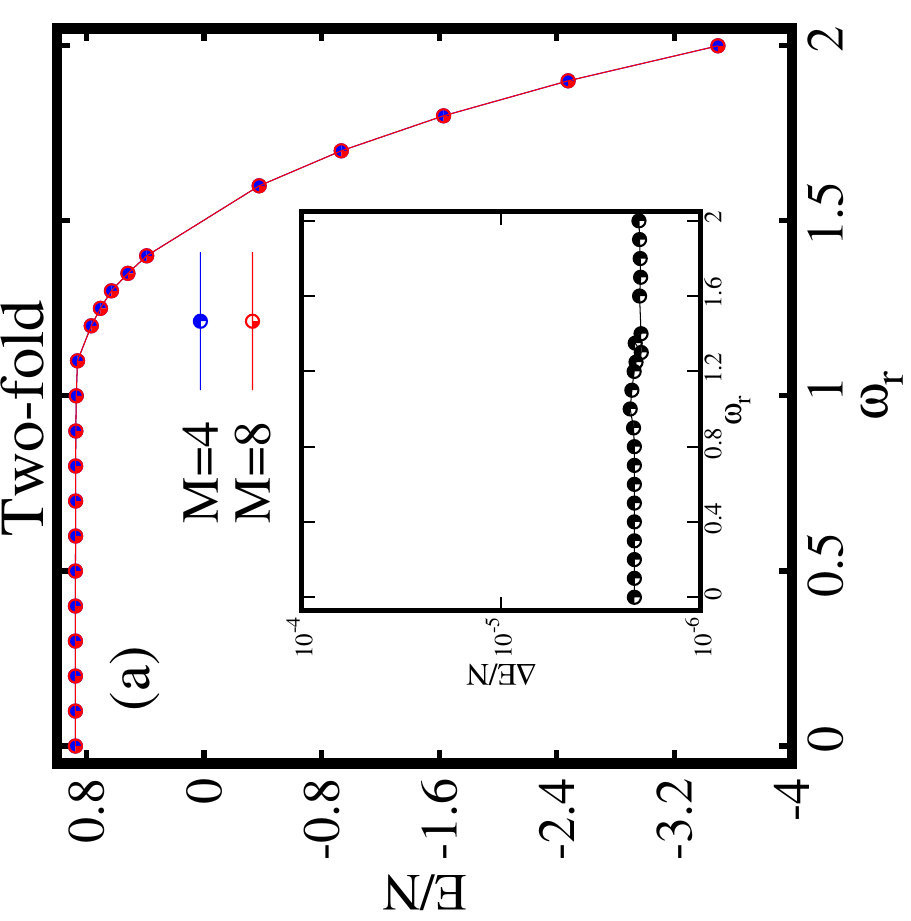}
  \includegraphics[angle=270,width=0.48\textwidth]{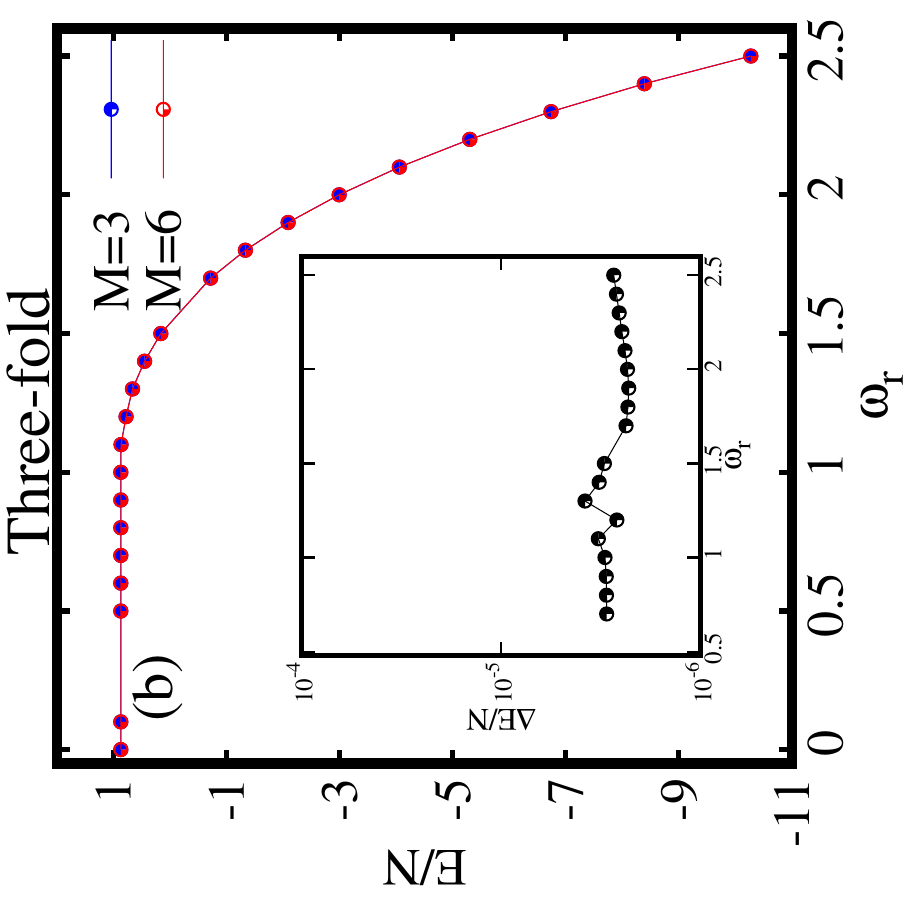}
  \includegraphics[angle=270,width=0.48\textwidth]{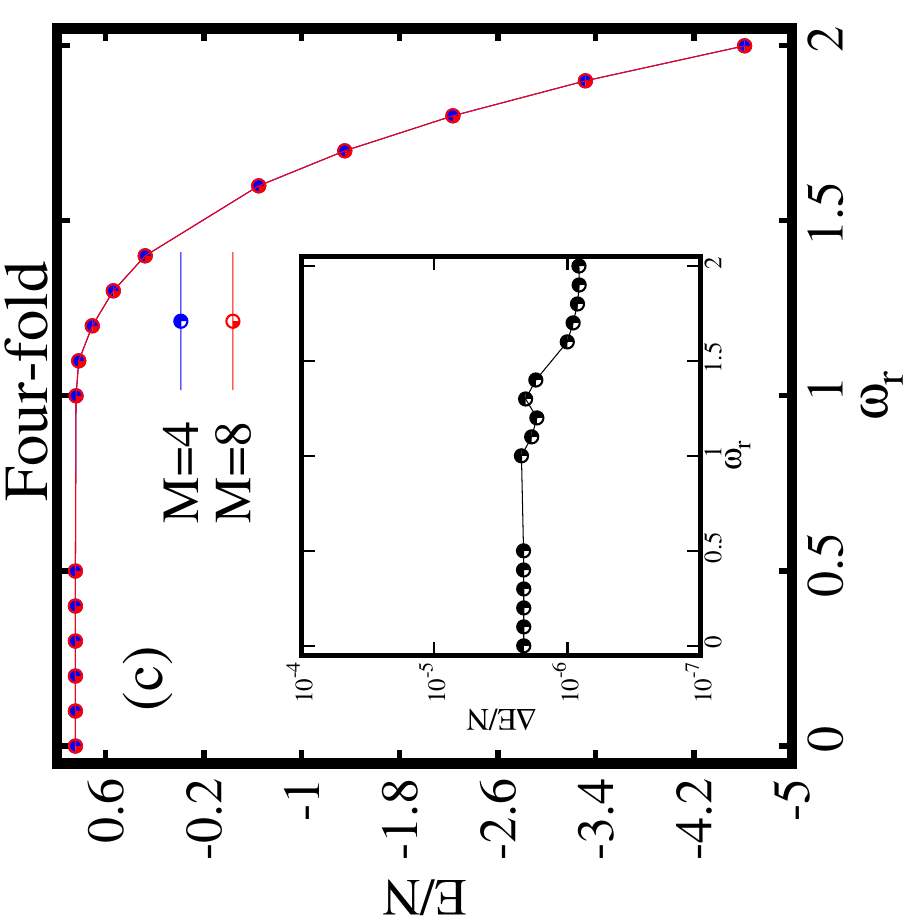}
 \caption{Many-body ground state energy per particle $\frac{E}{N}$ as a function of the rotation frequency $\omega_r$. (a) for the elongated trap $M=8$ self-consistent orbitals are used to show the convergence of the $M=4$ results, (b) for the three-fold symmetric trap $M=6$ self-consistent orbitals are used to check the convergence of $M=3$ results, and finally in (c) $M=8$ self-consistent orbitals are used to check the convergence of $M=4$ results of the four-fold symmetric trap. The inset shows the energy difference per particle as defined in Equation \eqref{energy_diff}. The quantities shown are dimensionless.}\label{Fig2}
\end{figure}
 namely the elongated trap
\begin{eqnarray}
 V(\mathbf{r})=\frac{1}{4}(0.8x^2+y^2)^2,
\label{eq:elongated}
\end{eqnarray}
the three-fold symmetric trap
\begin{eqnarray}
 V({\bf{r}}) = \frac{1}{5} (x^2+y^2)^2+\frac{1}{5}(x^2y-\frac{1}{3}y^3),
\label{eq:potentials_three}
\end{eqnarray}
and, finally, the four-fold symmetric trap defined as
\begin{eqnarray}
 V(\mathbf{r}) = \frac{1}{4}(x^4+y^4).   
\label{eq:potentials_four}
\end{eqnarray}
The convergence of the  many-body energies reported in the main text with $M$ self-consistent orbitals is here verified by taking $2M$ self-consistent orbitals and comparing the respective results.

Figure~\ref{Fig2} shows the behavior of the ground state energy $\frac{E}{N}$ as a function of the rotation frequency $\omega_r$ for different numbers of self-consistent orbitals for three confining potentials given by Equations \eqref{eq:elongated}, \eqref{eq:potentials_three}, and \eqref{eq:potentials_four}. The inset in all the three panels of Figure~\ref{Fig2} display the energy difference between the energies computed for two different orbital numbers $M$,
\begin{eqnarray}
\frac{\Delta E}{N}=\frac{E_{M}}{N}-\frac{E_{2M}}{N}.\label{energy_diff}
\end{eqnarray}

Figure \ref{Fig2}(a) corresponds to the elongated trap [Equation \eqref{eq:elongated}]. The convergence of energy is checked for two different orbitals $M=4,8$. We find that the two energy curves $\frac{E}{N}$ computed with $M=4,8$ orbitals fall on top of each other for all the frequencies. It is evident that our results converges
with respect to the number of orbitals for elongated trap. Similar convergence is observed in case of the three-fold symmetric trap with $M=3,6$ orbitals in Figure \ref{Fig2}(b) [see Equation \eqref{eq:potentials_three}] and the four-fold symmetric trap [Equation \eqref{eq:potentials_four}] with $M=4,8$ orbitals as evident from Figure \ref{Fig2}(c).
We conclude that the ground state energies are fully converged with respect to the number of orbitals in all the three confining potentials at all rotation frequencies.

\section{Convergence of the natural occupations}\label{sec:orbital_int}
\begin{figure}[htbp]
  \centering
  \includegraphics[angle=270,width=0.48\textwidth]{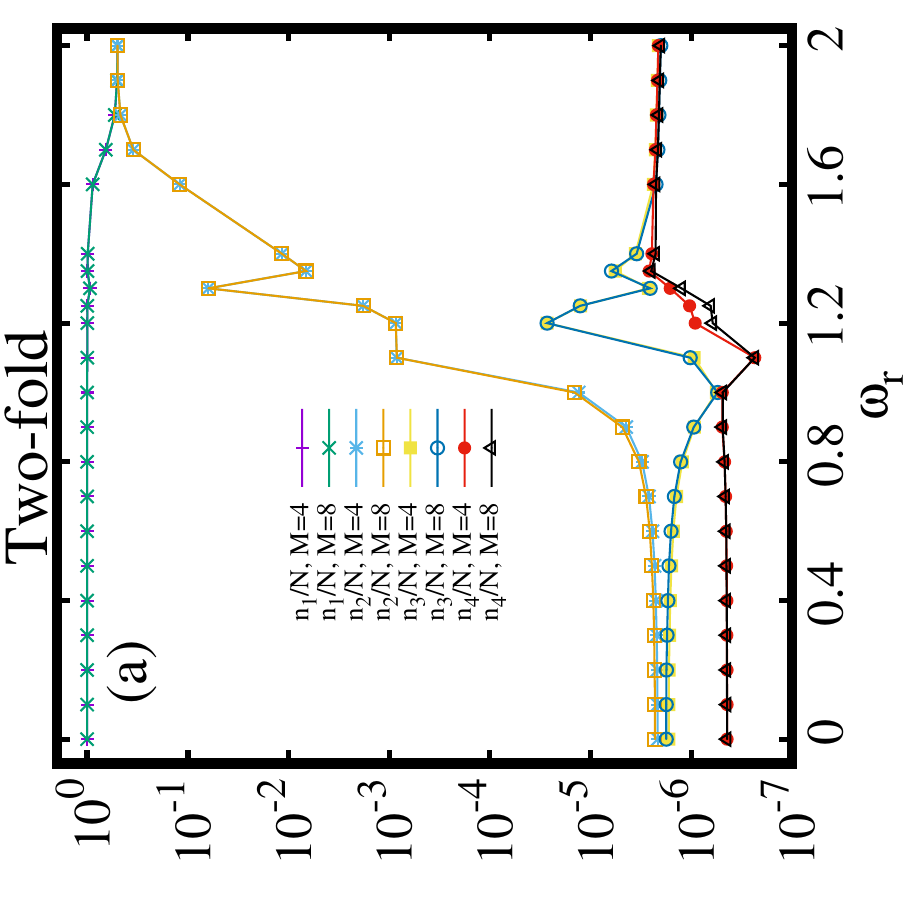}
   \includegraphics[angle=270,width=0.48\textwidth]{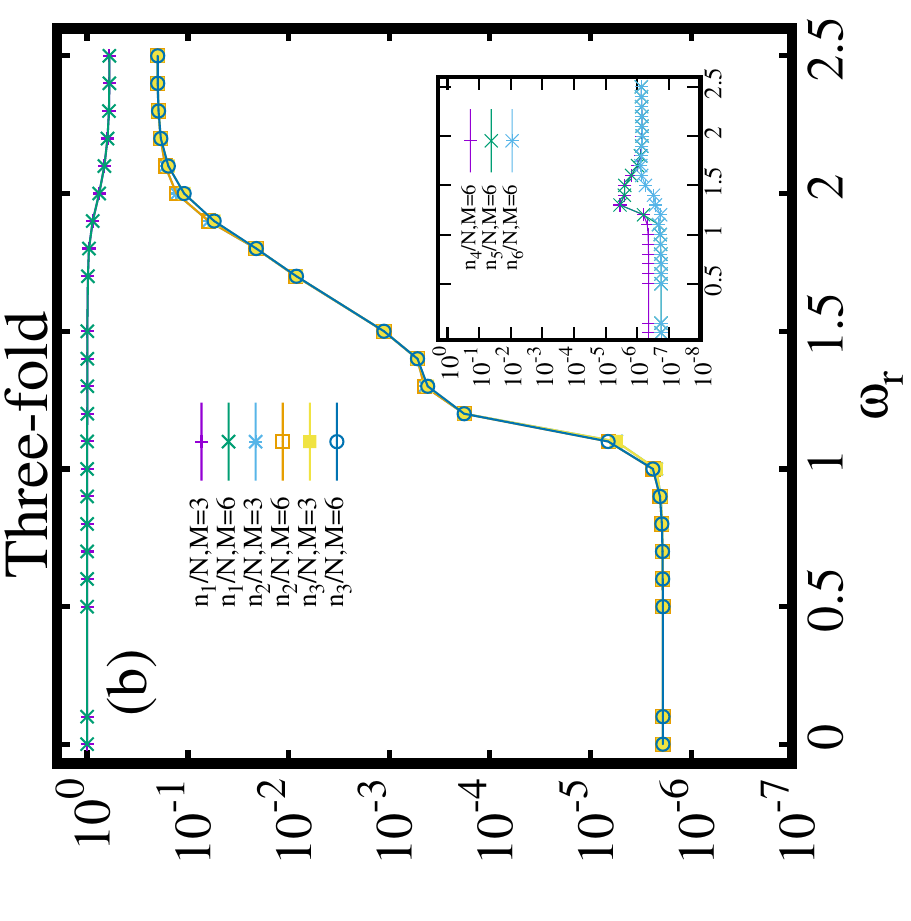}
   \includegraphics[angle=270,width=0.48\textwidth]{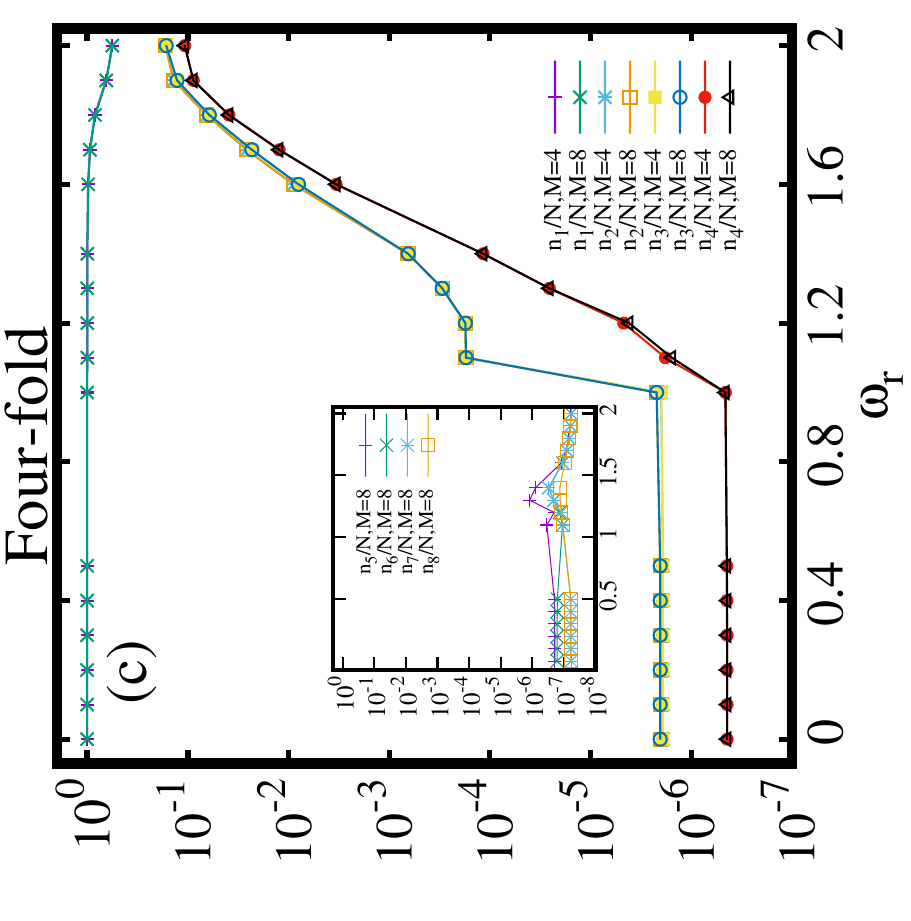}
 \caption{ Convergence of the natural occupations with respect to the number of natural orbitals. (a) for the elongated trap with M=4,8 self-consistent orbitals, (b) for the three-fold symmetric trap with M=3,6 self-consistent orbitals, and (c) for the four-fold symmetric trap with M=4,8 self-consistent orbitals. The inset in panels (b) and (c) depict the variation of the smallest natural occupations with rotation.}\label{Fig4}
 \end{figure}

The behavior of the natural occupations $n_{j}/N$ for the confining potentials defined in Equations \eqref{eq:elongated}, \eqref{eq:potentials_three}, and \eqref{eq:potentials_four}, as a function of the rotation frequency $\omega_r$
are shown in Figures~\ref{Fig4} for different orbital numbers $M$.
It is observed 
that all the natural occupations are fully converged for the three confining potentials. The inset in Figure \ref{Fig4}(b) shows the variation of the smallest natural occupations $\frac{n_4}{N}$, $\frac{n_5}{N}$, and $\frac{n_6}{N}$ with the rotation frequency $\omega_r$ for the three-fold symmetric trap. All the three natural occupations stays minimal with $\frac{n_4}{N}\sim \frac{n_5}{N}\sim \frac{n_6}{N}\sim 10^{-6}$ for all $\omega_r$. Similarly, the inset of Figure \ref{Fig4}(c) depicts the natural occupations $\frac{n_5}{N}\sim 10^{-6}$, $\frac{n_6}{N}\sim \frac{n_7}{N}\sim \frac{n_8}{N}\sim 10^{-7}-10^{-6}$ for the four-fold symmetric trap. The convergence of the natural occupations is evident in all three confining traps.

\section{Convergence of the position, momentum, and angular momentum variances}
The variance of a many-particle operator $\hat{A}=\sum_{j=1}^N \hat{a}_j$ per particle is defined as 
\begin{eqnarray}
 \frac{1}{N}\Delta^2_{\hat{A}}=\frac{1}{N}\bigg[\langle\Psi|\hat{A}^2|\Psi\rangle-\langle\Psi|\hat{A}|\Psi\rangle^2\bigg].
 \end{eqnarray}
 The detail discussion is included in the main text.
 Here, we report the convergence of many-particle variances of the position, momentum and angular momentum operators \cite{lode2012numerically,six} for all the three confining potentials in the rotating frame with respect to the number of orbitals.
 \begin{figure}[htbp]
  \centering
  \includegraphics[angle=270,width=0.45\textwidth]{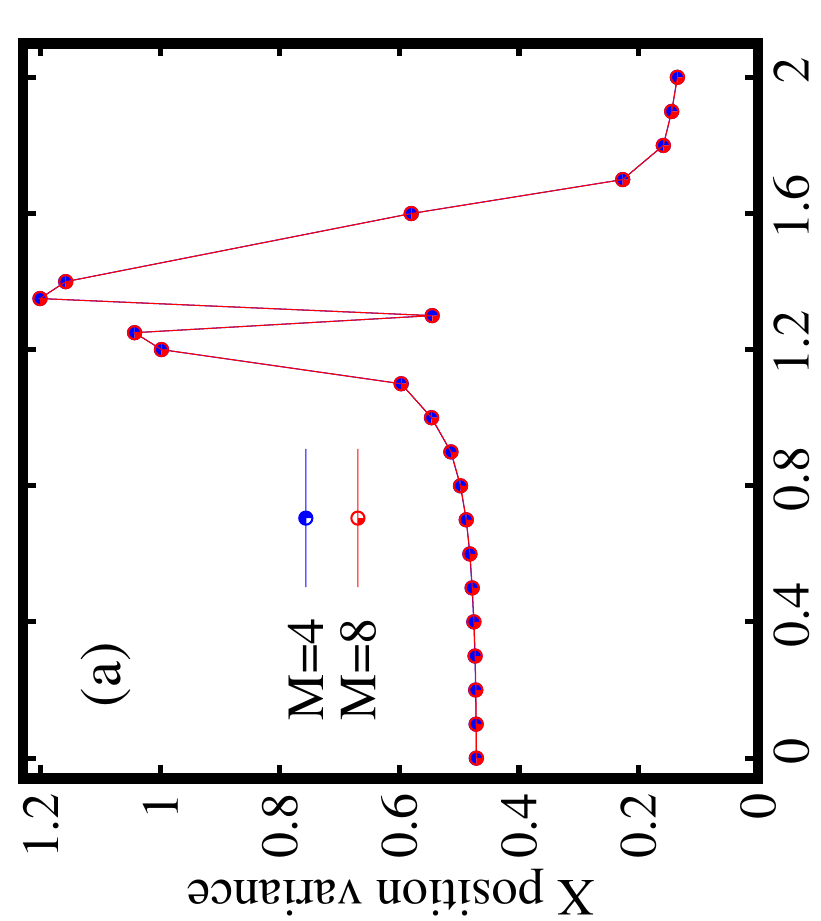}
  \includegraphics[angle=270,width=0.45\textwidth]{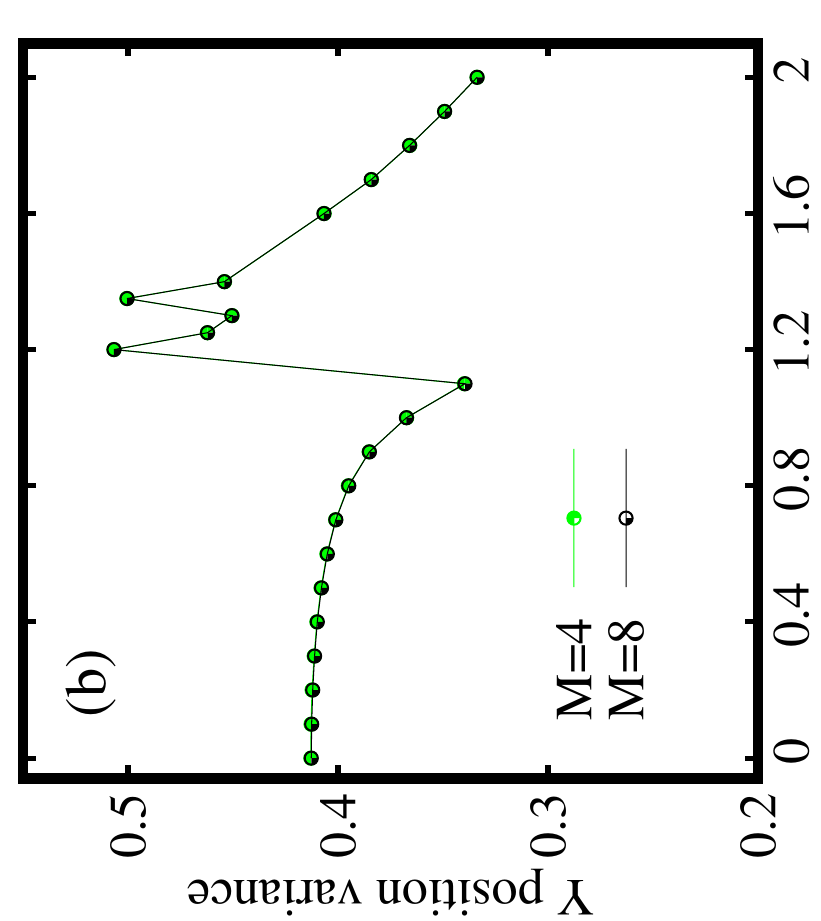}\\
  \includegraphics[angle=270,width=0.45\textwidth]{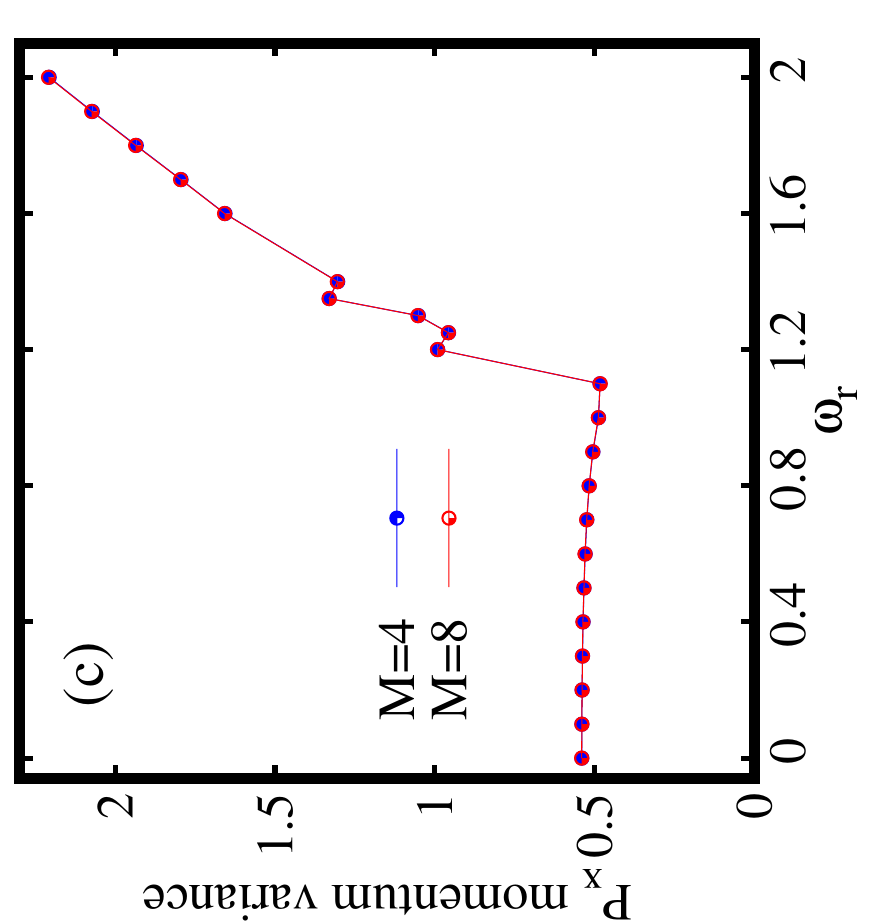}
  \includegraphics[angle=270,width=0.45\textwidth]{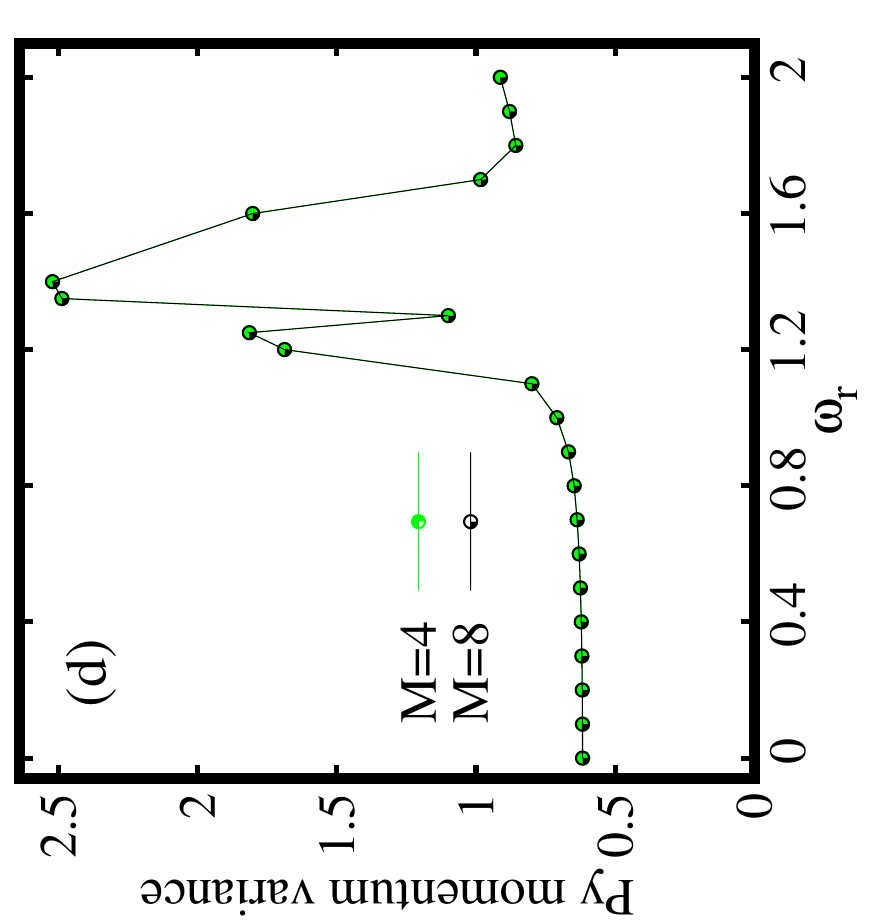}
 \caption{ Convergence of the many-particle position variances (a)-(b) and momentum variances (c)-(d) computed for the elongated trap with M=4,8 self-consistent orbitals along the $x$- and $y$- directions.}\label{po_variance_conv_elong}
 \end{figure}

\subsection{The position and momentum variances}\label{sec:Result}
 This section presents the convergence of many-particle position variances $\frac{1}{N}\Delta^2_{\hat{X}}$, $\frac{1}{N}\Delta^2_{\hat{Y}}$ and many-particle momentum variances $\frac{1}{N}\Delta^2_{\hat{P}_X}$, $\frac{1}{N}\Delta^2_{\hat{P}_Y}$ with respect to the number of orbitals $M$ for the three confining potentials. 
 
Figures \ref{po_variance_conv_elong}(a)-(b) show the convergence of the many-particle position variances and Figures \ref{po_variance_conv_elong}(c)-(d) correspond to the many-particle momentum variances for the elongated trap [Equation \eqref{eq:elongated}] with $M=4,8$ orbitals.

\begin{figure}[htbp]
  \centering
  \includegraphics[angle=270,width=0.45\textwidth]{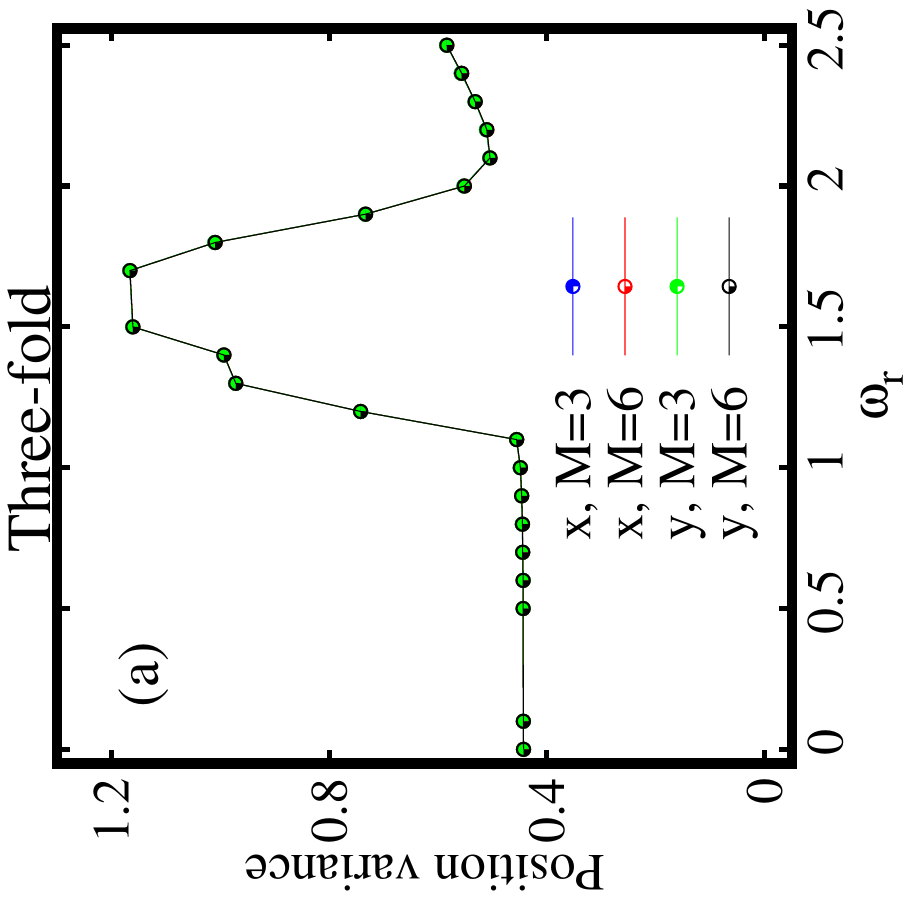}
  \includegraphics[angle=270,width=0.45\textwidth]{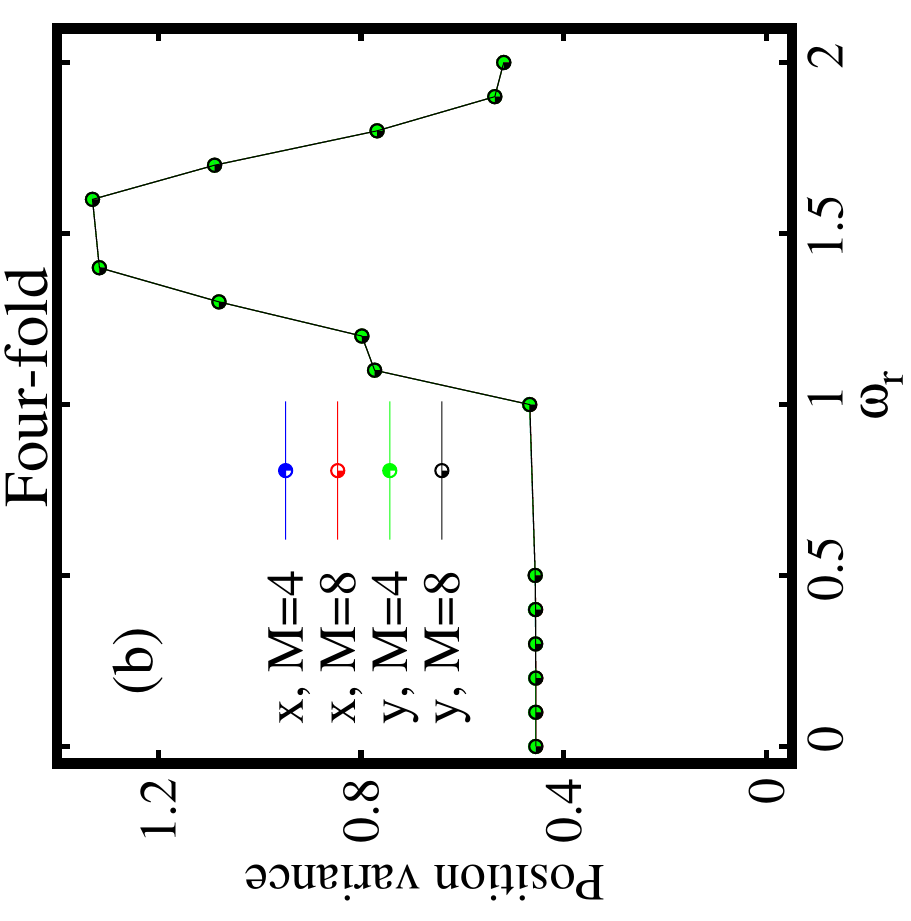}\\
  \includegraphics[angle=270,width=0.45\textwidth]{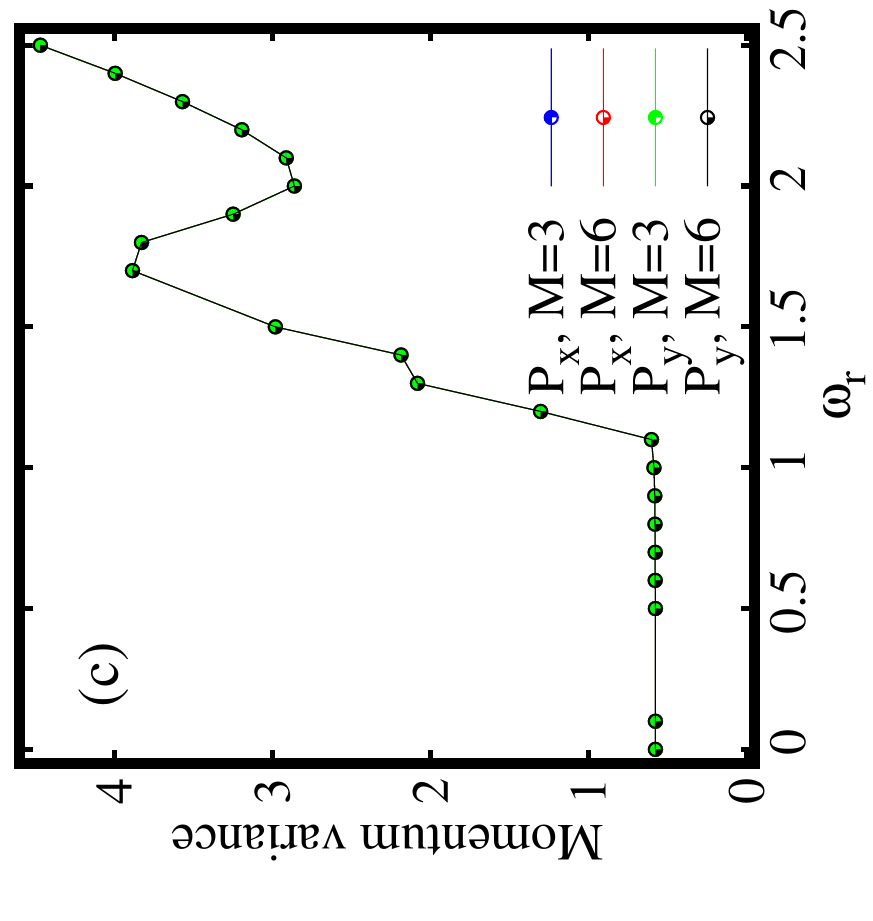}
  \includegraphics[angle=270,width=0.45\textwidth]{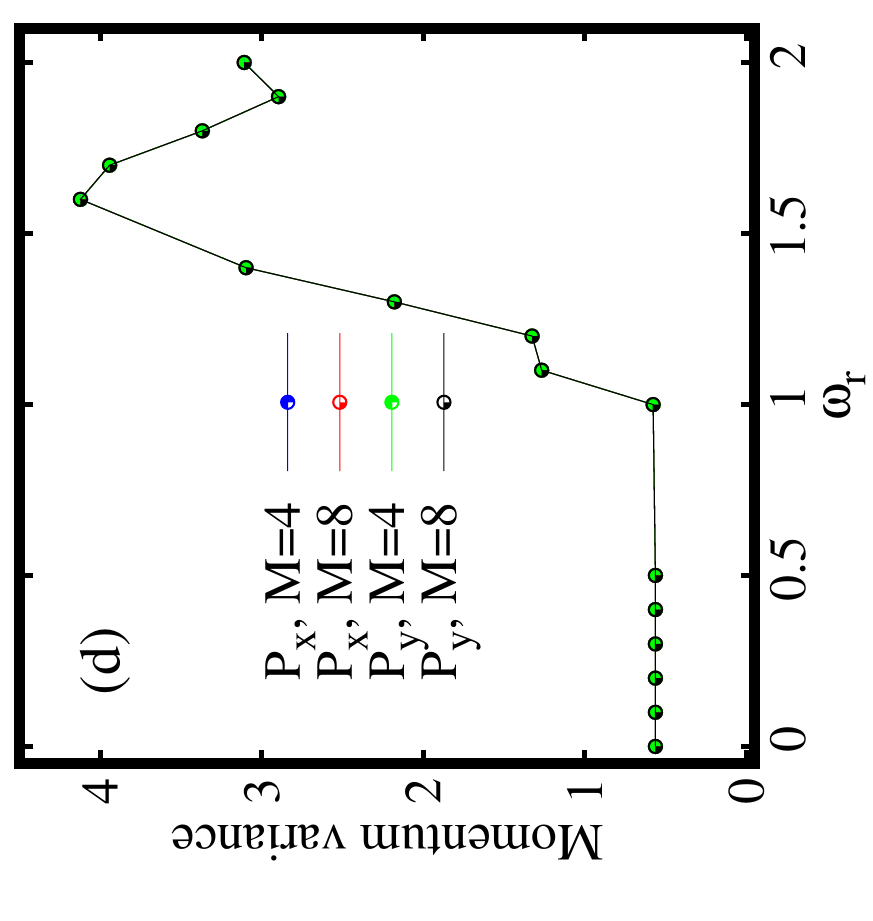}
 \caption{Convergence of the many-particle position (a) and momentum variances (c) with M=3,6 self-consistent orbitals for the three-fold symmetric trap. Panels (b) and (d) correspond to many-particle position and momentum variances with M=4,8 self-consistent orbitals for the four-fold symmetric trap.}\label{mo_variance_conv_three}
 \end{figure}
Figures \ref{mo_variance_conv_three}(a),(c) correspond to the many-particle position and momentum variances for three-fold symmetric trap [Equation \eqref{eq:potentials_three}]. Similarly, Figures \ref{mo_variance_conv_three}(b),(d) 
depict the many-particle position and momentum variances for the four-fold symmetric trap [Equation \eqref{eq:potentials_four}]. 

It is evident that the convergence with respect to the number orbitals of the many-particle variances of the position and momentum operators for all the three confining potentials is excellent.
 \subsection{Angular momentum properties}
 Before we discuss the angular momentum variance, we wish to show that the expectation value of the angular momentum operator discussed in the main text converges at the many-body level. Then, we proceed to the angular momentum variance, and augment the text by its analysis.
 \begin{figure}[htbp]
  \centering
  \includegraphics[angle=270,width=0.327\textwidth]{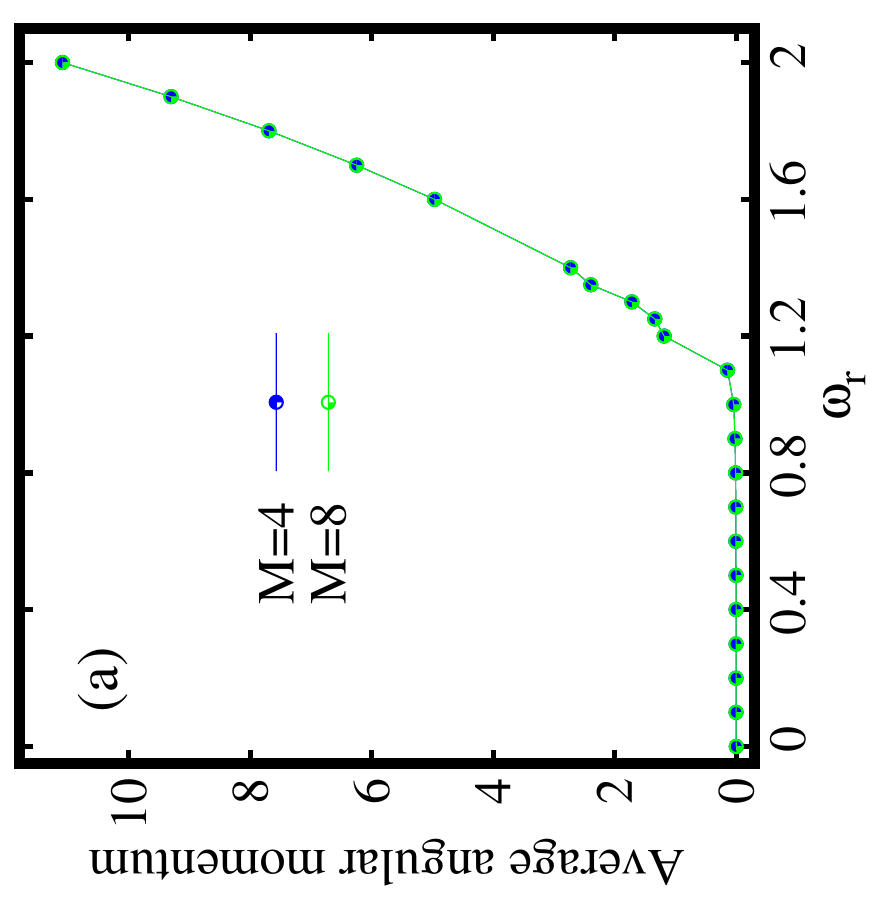}
  \includegraphics[angle=270,width=0.327\textwidth]{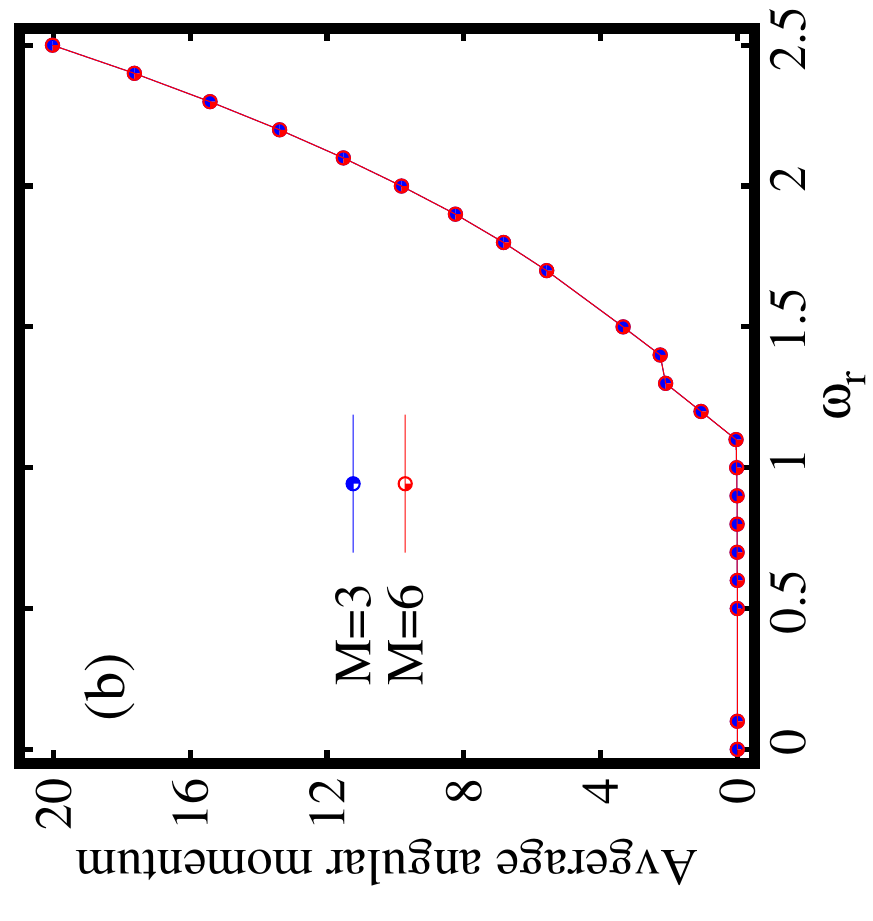}
  \includegraphics[angle=270,width=0.327\textwidth]{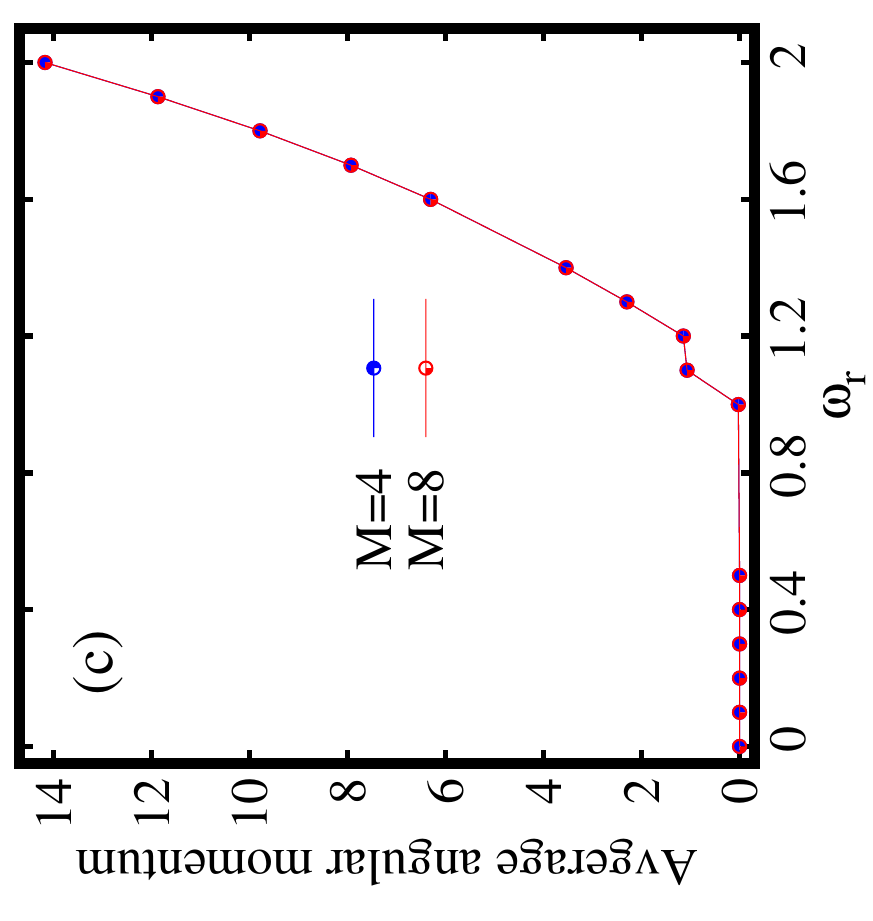}
 \caption{ Convergence of the expectation values of the angular momentum operator per particle. (a) for the elongated trap with M=4,8 self-consistent orbitals, (b) for the three-fold symmetric trap with M=3,6  self-consistent orbitals and (c) for the four-fold symmetric trap with M=4,8  self-consistent orbitals.}\label{ang_mom_conv_elong}
 \end{figure}

 Figure (\ref{ang_mom_conv_elong}) proves the convergence of the expectation values of angular momentum operator $\hat{L}_Z=\sum_{j=1}^N\hat{l}_{z_j}$ per particle $\frac{1}{N}\langle \Psi|\hat{L}_Z|\Psi\rangle$, with respect to the number of orbitals $M$ for the three confining
 \begin{figure}[htbp]
\centering
\includegraphics[angle=270,width=0.327\textwidth]{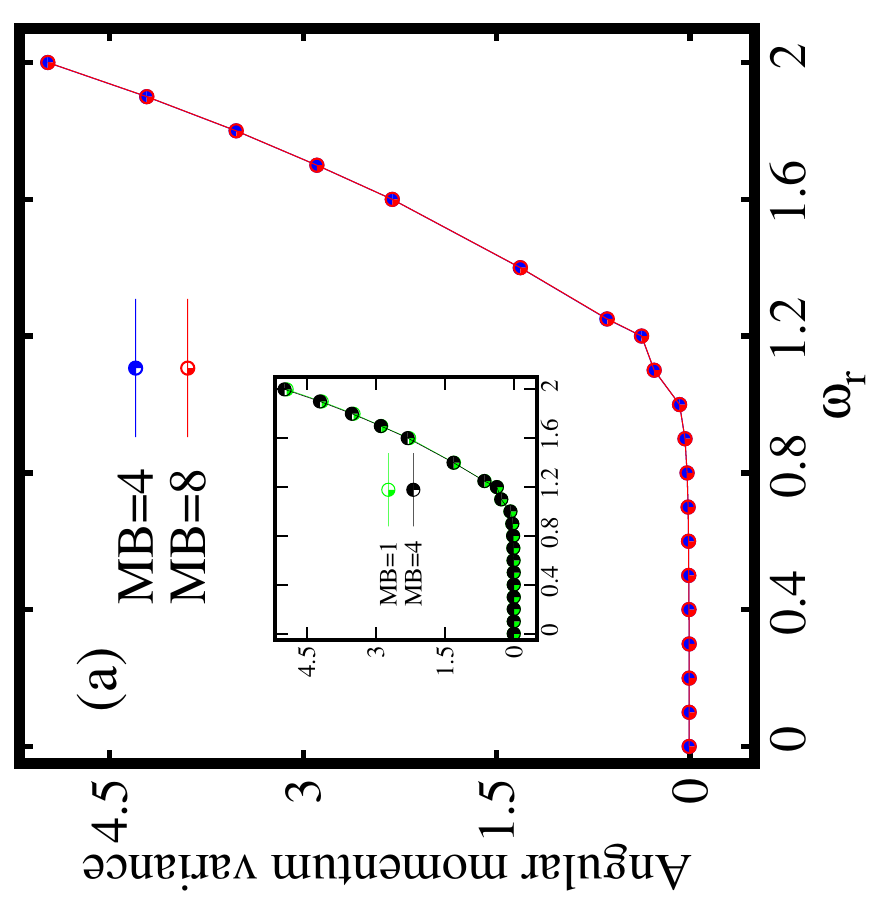}
\includegraphics[angle=270,width=0.327\textwidth]{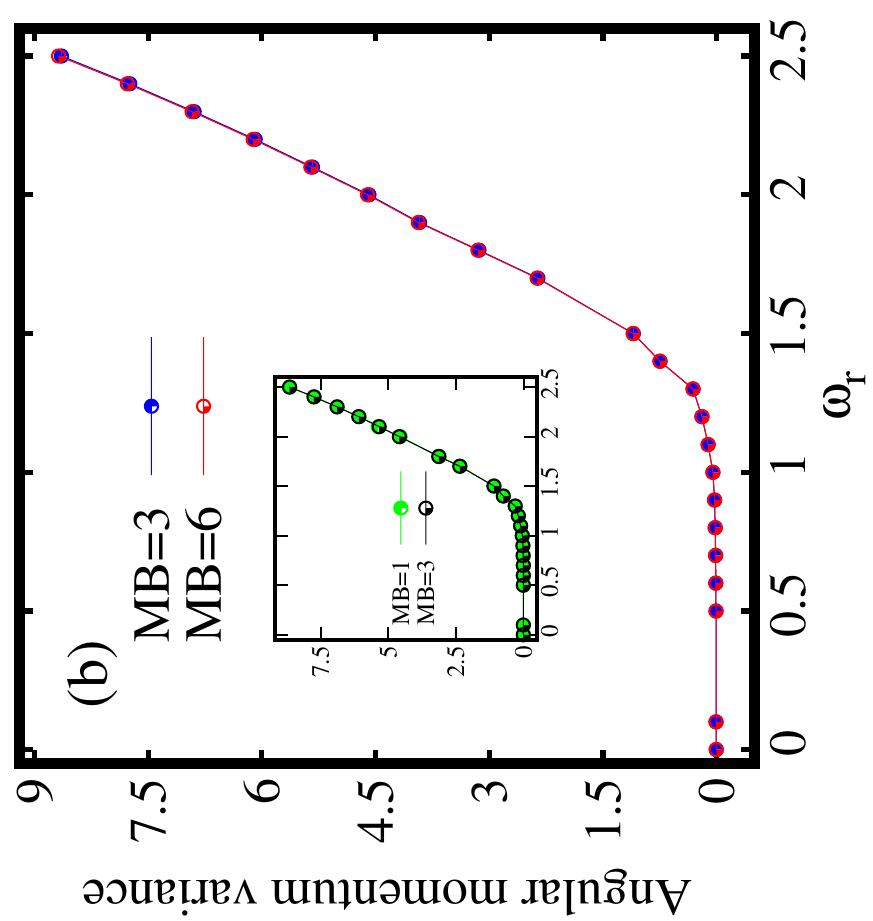}
\includegraphics[angle=270,width=0.327\textwidth]{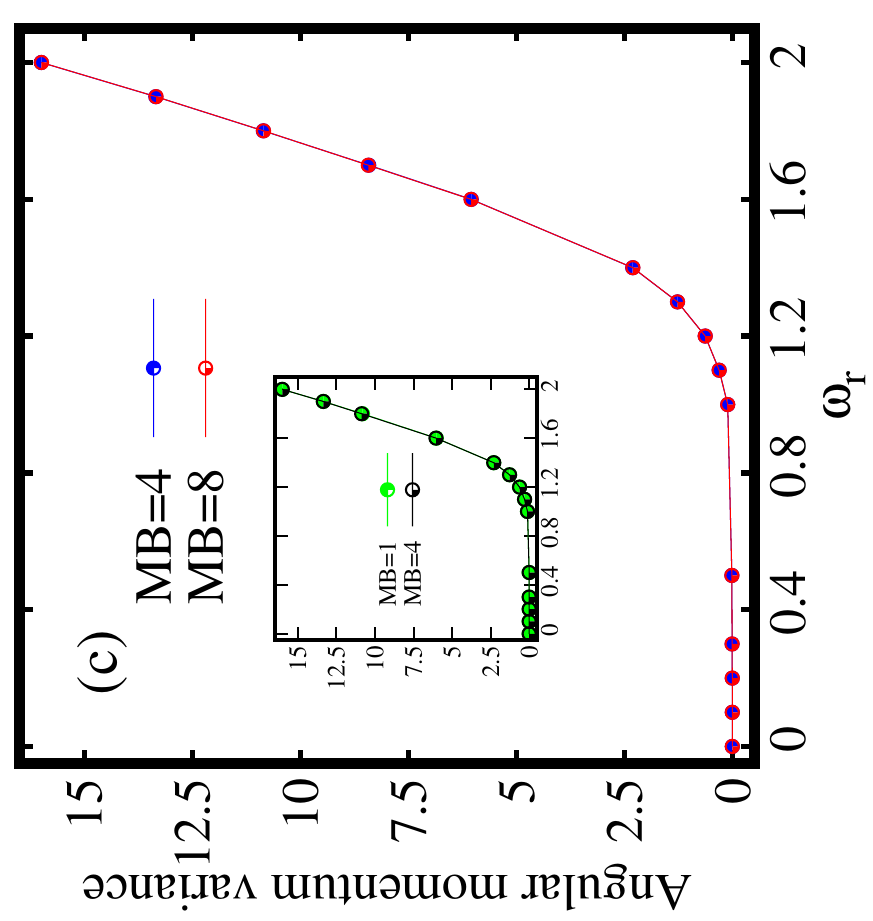}
\caption{Convergence of the angular momentum variance. (a) for the elongated trap with M=4,8 self-consistent orbitals, (b) for the three-fold symmetric trap with M=3,6 self-consistent orbitals and (c) for the four-fold symmetric trap with M=4,8 self-consistent orbitals. The insets of (a), (b), and (c) demonstrate the angular momentum variances at the mean-field and many-body levels that are essentially the same in our rotating traps.}
\label{var_lz}
\end{figure}
 potentials; the elongated trap [Equation \eqref{eq:elongated}], the three-fold symmetric trap [Equation \eqref{eq:potentials_three}] and the four-fold symmetric trap [Equation \eqref{eq:potentials_four}].

Finally, we discuss the convergence of the many-particle angular momentum 
variance $\frac{1}{N}\Delta^2_{\hat{L}_Z}$ in all the three confining traps. It is evident from Figure (\ref{var_lz}) that the angular momentum variances are fully converged with respect to the number of orbitals for all the three confining potentials at all $\omega_r$. The insets of Figure (\ref{var_lz}) demonstrate that in the present rotating traps and interaction strengths the mean-field and many-body angular momentum variances essentially coincide. 

\end{document}